\documentclass[12pt]{elsarticle}

\usepackage[american]{babel}
\usepackage[utf8]{inputenc}
\usepackage{ae,aecompl}
\usepackage{booktabs}
\usepackage{subfigure}

\usepackage{amsmath} 
\usepackage{amssymb}
\usepackage{commath}
\usepackage{dsfont}

\everymath={\displaystyle}

\renewcommand{\vec}[1]{\ensuremath{\mathbf{#1}}}

\newcommand{\mtx}[1]{\ensuremath{\left[#1\right]}}

\newcommand{\massop}[2]{\ensuremath{ \int_{0}^{L} \rho \, A \, \ddot{#1} (x,t) \, #2(x) \, dx + m \, \ddot{#1} (L,t) \, #2(L) }}
\newcommand{\M}{\ensuremath{\mathcal{M}}}

\newcommand{\amassop}[2]{\ensuremath{ \int_{0}^{L} \rho \, A \, #1 (x,t) \, #2(x) \, dx }}
\newcommand{\aM}{\ensuremath{\widetilde{\mathcal{M}}}}

\newcommand{\dampop}[2]{\ensuremath{ \int_{0}^{L} c \, \dot{#1} (x,t) \, #2(x) \, dx }}
\newcommand{\C}{\ensuremath{\mathcal{C}}}

\newcommand{\stiffop}[2]{\ensuremath{ \int_{0}^{L} E \, A \, #1' (x,t) \, #2' (x) \, dx+ k \, #1(L,t) \, #2(L) }}
\newcommand{\K}{\ensuremath{\mathcal{K}}}

\newcommand{\forceop}[1]{\ensuremath{ \int_{0}^{L} f(x,t) \, #1 (x) \, dx }}
\newcommand{\F}{\ensuremath{\mathcal{F}}}

\newcommand{\nlforceop}[2]{\ensuremath{ -k_{NL} \, \left( u(L,t) \right)^3 #2 (L) }}
\newcommand{\NL}{\ensuremath{\mathcal{F}_{NL}}}

\newcommand{\indfunc}[1]{\ensuremath{\mathds{1}_{#1} }}

\newcommand{\dimless}[1]{\ensuremath{ #1^{\ast} }}

\newcommand{\R}{\ensuremath{\mathbb{R}}}           
\newcommand{\U}{\ensuremath{\mathcal{U} }}   
\newcommand{\W}{\ensuremath{\mathcal{W} }}   

\renewcommand{\SS}[1][]{\ensuremath{\Theta^{#1} }}

\newcommand{\SSpt}[1][]{\ensuremath{\theta}}

\newcommand{\SF}[1][]{\ensuremath{{\mathbb{A}^{#1}} }}

\newcommand{\PM}[1][]{\ensuremath{{\mathbb{P}^{#1}} }}

\newcommand{\randvar}[1]{\ensuremath{#1}}

\newcommand{\randproc}[1]{\ensuremath{#1}}

\newcommand{\pdf}[1]{\ensuremath{p_{\tiny{#1}}}}

\newcommand{\entropy}[1]{\ensuremath{\mathbb{S}\left[  #1 \right] }}

\newcommand{\entropyop}[3]{\ensuremath{ - \int_{#2}^{#3} #1(\xi) \, \ln \left( #1(\xi) \right) \, d\xi}}

\newcommand{\expvalop}[3]{\ensuremath{\int_{#2}^{#3} #1(\xi) \, \pdf{#1}(\xi) \, d\xi} }

\begin{document}

\begin{frontmatter}

\journal{Applied Mathematical Modelling}

\title{On the nonlinear stochastic dynamics of a continuous system with discrete attached elements}

\author[pucmec]{Americo Cunha Jr\corref{cor}}
\ead{americo.cunhajr@gmail.com}
\author[pucmec]{Rubens Sampaio}
\ead{rsampaio@puc-rio.br}

\cortext[cor]{Corresponding author.}

\address[pucmec]{Department of Mechanical Engineering, PUC--Rio\\
Rua Marqu\^{e}s de S\~{a}o Vicente, 225, G\'{a}vea, Rio de Janeiro - RJ, Brazil - 22453-900}

\begin{abstract}
This paper presents a theoretical study on the influence of a discrete element 
in the nonlinear dynamics of a continuous mechanical system subject to 
randomness in the model parameters. This system is composed by an
elastic bar, attached to springs and a lumped mass, with a random elastic
modulus and subjected to a Gaussian white-noise distributed external force.
One can note that the dynamic behavior of the bar is significantly altered 
when the lumped mass is varied, becoming, on the right extreme and  
for large values of the concentrated mass, similar to a mass-spring system.
It is also observed that the system response is more influenced by the 
randomness for small values of the lumped mass. The study conducted 
also show an irregular distribution of energy through the spectrum 
of frequencies, asymmetries and multimodal behavior in the 
probability distributions of the lumped mass velocity.
\end{abstract}


\end{frontmatter}

\pagebreak
\section{Introduction}

A couple of engineering structure has small parts whose dimensions are 
negligible compared to the entire structure, but its presence induces significant 
effects on its behavior. In this situation it is common to model the structure as a 
continuous system with discrete elements coupled. The open literature reports 
studies that use such continuous/discrete models for the analysis of
drillstrings \cite{ritto2013p145}, 
carbon nanotubes \cite{aydogdu2011p1229,murmu2011p62},
naval structure-motor coupling \cite{rossit2001p933},
beams coupled with springs \cite{gurgoze2011p239,mao2011p756},
a damper \cite{hagedorn1987p352} and/or a discrete mass 
\cite{andrews2002p1033}, etc.

Like any computational model, these continuous/discrete models are
subjected to uncertainties. These uncertainties are due to the variability of 
the model parameters (physical constants, geometry, etc), and mainly due to 
the possible inaccuracies committed in the model conception 
(wrong hypotheses about the physics)
\citep{soize2000p277, soize2005p1333,soize2013p2379}.

In~this~sense, this work intends to analyze the influence of discrete elements
in a continuous mechanical system subjected to randomness in the model parameters.
For this, it is considered a one-dimensional elastic bar, with random elastic modulus, 
fixed on the left extreme and with a lumped mass and two springs (one linear and another nonlinear) 
on the right extreme, with viscous damping, and subjected to an external force which 
is proportional to a Gaussian white-noise. The theoretical study developed aims to 
illustrate a consistent methodology to analyze the influence of coupled discrete elements 
into the stochastic dynamics of nonlinear mechanical systems.
The results of this study complement a series of preliminary studies made on the same
subject \citep{cunhajr2012p2673,cunhajr2012uncert,cunhajr2013seeccm,cunhajr2013diname}.

The work is organized as follows. The section~\ref{determ_approach}
presents the deterministic equation of the model, its variational form,
and the discretization procedure used to solve it. The stochastic modeling 
of the problem is shown in section~\ref{stochastic_approach}, as well as 
the construction of a probability distribution for the elastic modulus, 
using the maximum entropy principle.  In the section~\ref{num_experim}, 
some configurations of the model are analyzed in order to characterize the 
effect of lumped mass in the nonlinear dynamical system. Finally, 
in the section~\ref{concl_remaks}, the main conclusions are emphasized.


\section{Deterministic modeling of the mechanical system}
\label{determ_approach}

\subsection{Strong form of the initial--boundary value problem}
\label{strong_form}

\begin{figure}[h!]
	\centering
	\includegraphics[scale=0.6]{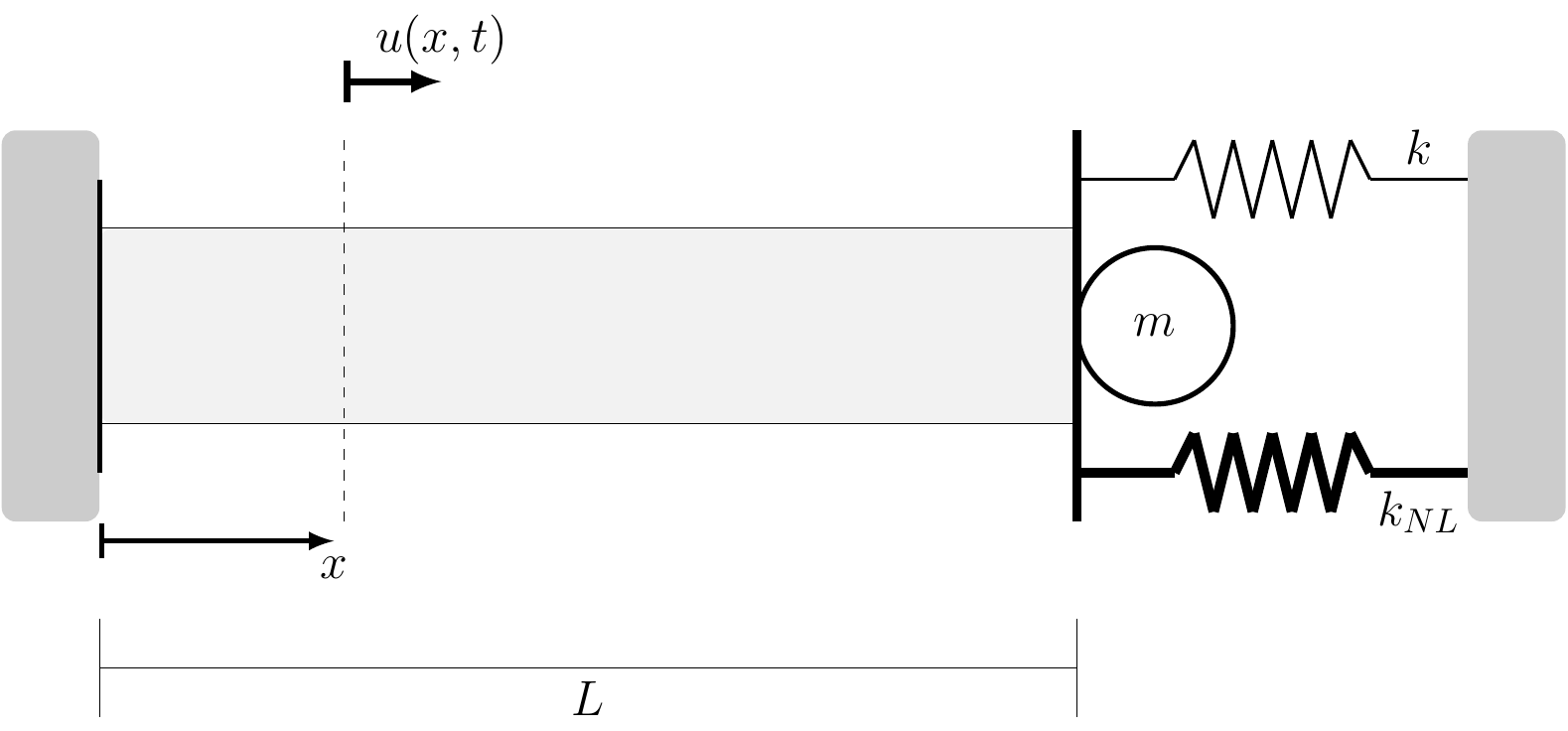}
	\caption{Sketch of a bar fixed at one end, and attached
					to two springs and a lumped mass on the other extreme.}
	\label{bar_fig}
\end{figure}

The mechanical system that will be studied in this work is presented
Figure~\ref{bar_fig}. It consists of an elastic bar for which the left side is fixed 
at a rigid wall, and the right side is attached to a lumped mass and two springs 
(one linear and one nonlinear). For simplicity, from now on, this system will
be called the fixed-mass-spring bar or simply the bar.

The bar displacement field\footnote{The field $u$ is implicitly assumed 
to be as regular as needed for the initial--boundary value problem of 
Eqs.(\ref{bar_eq}) to (\ref{bar_ic}) to be well posed.} $u$,
which depends on the position $x$ and the time $t$, evolves, 
for all $(x,t) \in (0,L) \times (t_0,t_f]$, according to

\begin{eqnarray}
    \rho A \dpd[2]{u}{t} + c \dpd{u}{t}  
    & = & \dpd{}{x}\left( E A \dpd{u}{x}  \right) \\ \nonumber
    & - & \left( k u + k_{NL} u^3 + m \dpd[2]{u}{t} \right) \delta(x-L)
    + f(x,t),
    \label{bar_eq}
\end{eqnarray}

\noindent
where $\rho$ is mass density, $A$ is the cross section area,
$c$ is the damping coefficient, $E$ is the elastic modulus, 
$k$ is the stiffness of the linear spring, $k_{NL}$ is the stiffness 
of the nonlinear spring, $m$ is the lumped mass, and $f$ is a 
distributed external force, which depends on $x$ and $t$.
The symbol $\delta(x-L)$ denotes the delta of Dirac distribution
at $x=L$, where $L$ is the bar length.

The boundary conditions for this problem are given by

\begin{equation}
    u(0,t) = 0,
    ~~\mbox{and}~~
    E A \dpd{u}{x} (L,t) = 0,
    \label{bar_bc}
\end{equation}

\noindent
and the initial position and the initial velocity of the bar are

\begin{equation}
    u(x,t_0) = u_0(x),
    ~~\mbox{and}~~
    \dpd{u}{t}(x,t_0) = v_0(x),
    \label{bar_ic}
\end{equation}

\begin{figure}[h!]
		\centering
		\includegraphics[scale=1]{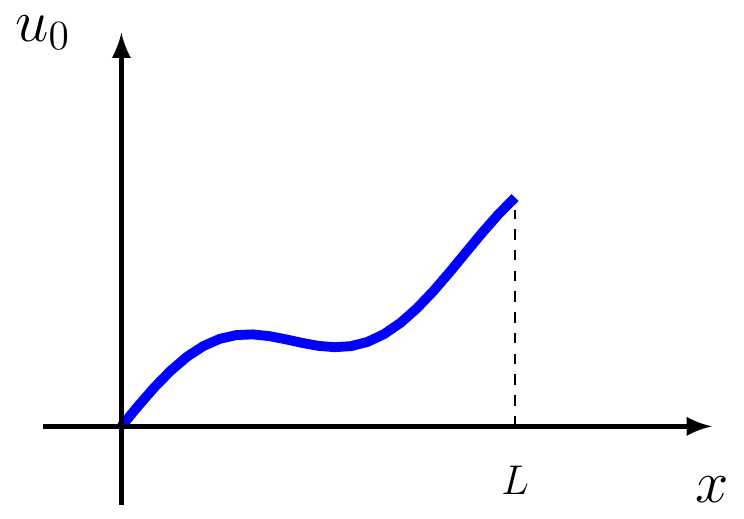}
		\caption{This figure illustrates the graph of $u_0$, the initial displacement of the bar.}
		\label{initial_disp_fig}
\end{figure}

\noindent
$u_0$ and $v_0$ being known functions of $x$, defined for $0 \leq x \leq L$.
For instance, 

\begin{equation}
    u_0(x) = \alpha_1 \phi_3(x) + \alpha_2 x, 
    ~~\mbox{and}~~ 
    v_0(x) = 0,
\end{equation}

\noindent
where $\alpha_1$ and $\alpha_2$ are constants, and 
$\phi_3$ is the third mode\footnote{Further details in 
the section~\ref{modes_nat_freq} } of the bar.
Note that $u_0$ reaches the maximum value at $x = L$,
see Figure~\ref{initial_disp_fig} for instance. 
This function is used to ``activate" the spring cubic nonlinearity, 
which depends on the displacement at $x = L$.

\subsection{Weak form of the initial--boundary value problem}
\label{weak_form}

Let $\U_t$ be a class of (time dependent) basis functions
and $\W$ be a class of weight functions. These sets are chosen 
as the space of functions with square integrable spatial derivative, 
which satisfy the essential boundary condition defined by Eq.(\ref{bar_bc}).

The weak formulation of the initial--boundary value problem above consists in finding,
for all $w$ in $\W$, a displacement field $u$ in $\U_t$ such that the following equations 
are satisfied

\begin{equation}
    \M(\ddot{u},w) + \C(\dot{u},w) + \K(u,w) = \F(w) + \NL(u,w),
    \label{weak_eq}
\end{equation}

\begin{equation}
	\aM(u(\cdot,t_0),w) = \aM(u_0,w),
    \label{weak_ic_eq1}
\end{equation}

\noindent
and

\begin{equation}
	\aM(\dot{u}(\cdot,t_0),w) = \aM(v_0,w),
    \label{weak_ic_eq2}
\end{equation}

\noindent
where 
$\M$ is the mass operator,
$\C$ is the damping operator,
$\K$ is the stiffness operator, 
$\F$ is the distributed external force operator,
$\NL$ is the nonlinear force operator, and
$\aM$ is the associated mass operator.
These operators are, respectively, defined as

\begin{eqnarray}
    \M(\ddot{u},w) & = & \massop{u}{w}, \label{mass_op} \\
    \C(\dot{u},w) & = &\dampop{u}{w}, \label{damp_op} \\
    \K(u,w) & = & \stiffop{u}{w}, \label{stiff_op} \\
    \F(w) & = & \forceop{w}, \label{force_op}\\
    \NL(u,w) & = & \nlforceop{u}{w}, \label{nlforce_op}\\
    \aM(u,w) & = & \amassop{u}{w}, \label{amass_op}
\end{eqnarray}

\noindent
where ~$\dot{}$~ is an abbreviation for time derivative and
~$'$~ is an abbreviation for spatial derivative.

\subsection{Linear conservative dynamics associated}
\label{eigen_probl}

Consider the linear homogeneous equation associated 
to the Eq.(\ref{weak_eq}),

\begin{equation}
    \M(u,w) + \K(u,w)  = 0,
    \label{homog_weak_eq}
\end{equation}

\noindent
obtained when disposing the dissipation and the external forces
acting on the mechanical system.

Assume that Eq.(\ref{homog_weak_eq}) has a solution of the form
$u(x,t) = e^{i \nu t} \phi(x)$, where $\nu$ is the
natural frequency, $\phi$ is mode and
$i=\sqrt{-1}$ is the imaginary unit.
Replacing this expression of $u$ in the Eq.(\ref{homog_weak_eq}),
and using the linearity of the operators $\M$ and $\K$, one gets

\begin{equation}
		\left(  - \nu^2 \M(\phi,w) + \K(\phi,w) \right)  e^{i \nu t} = 0,
\end{equation}

\noindent
which, since $e^{i \nu t} \neq 0$ for all $t$, is equivalent to

\begin{equation}
		- \nu^2 \M(\phi,w) + \K(\phi,w) = 0,
		\label{gen_eig_eq}
\end{equation}

\noindent
a generalized eigenvalue problem.

In order to solve the generalized eigenvalue problem defined by
Eq.(\ref{gen_eig_eq}), the technique of separation of variables 
is employed, which leads to a Sturm-Liouville problem \citep{algwaiz2007},
with denumerable number of solutions.
Therefore, this problem has a denumerable number of solutions,
all of then such as the following eigenpair
$(\nu_n^2,\phi_n)$, where $\nu_n$ is the
$n$-th natural frequency and
$\phi_n$ is the $n$-th mode of the system.

Note that, the eigenfunctions
$\{\phi_n\}_{n=1}^{+\infty}$ span the space of 
functions which contains the solution of the 
Eq.(\ref{gen_eig_eq})  \citep{brezis2010}.
Also, as can be seen in \cite{hagedorn2007}, 
these eigenfunctions satisfy, for all  $m \neq n$,
the orthogonality relations given by

\begin{equation}
		\M(\phi_n,\phi_m) = 0,
		\label{modalM_eq}
\end{equation}

\noindent
and

\begin{equation}
		\K(\phi_n,\phi_m) = 0.
		\label{modalK_eq}
\end{equation}

The characteristics listed above made the modes of the system 
good choices for the basis function, when one uses a weighted 
residual procedure \citep{scriven1966p735} to approximate 
the solution of the nonlinear variational problem defined by
Eqs.(\ref{weak_eq}) to (\ref{weak_ic_eq2}).

\subsection{Modes and natural frequencies}
\label{modes_nat_freq}

According to \cite{blevins1993}, a fixed-mass-spring bar 
has its natural frequencies and the corresponding orthogonal modes shape given by

\begin{equation}
		\nu_n = \lambda_n \frac{\bar{c}}{L},
		\label{nat_freq_eq}
\end{equation}

\noindent
and

\begin{equation}
		\phi_n (x) = \sin{ \left( \lambda_n \frac{x}{L} \right) },
		\label{shape_mod_eq}
\end{equation}

\noindent
where $\bar{c} = \sqrt{E/\rho}$ is the wave speed, and the 
$\lambda_n$ are the solutions of

\begin{equation}
		\cot{ \left( \lambda_n \right) }  + 
		\left( \frac{kL}{AE} \right) \frac{1}{\lambda_n} -
		\left( \frac{m}{\rho AL} \right) \lambda_n = 0.
		\label{lambda_eq}
\end{equation}

\begin{figure} [h!]
				\centering
				\subfigure[nat. freq. $\approx 1.15~kHz$]{
				\includegraphics[scale=0.28]{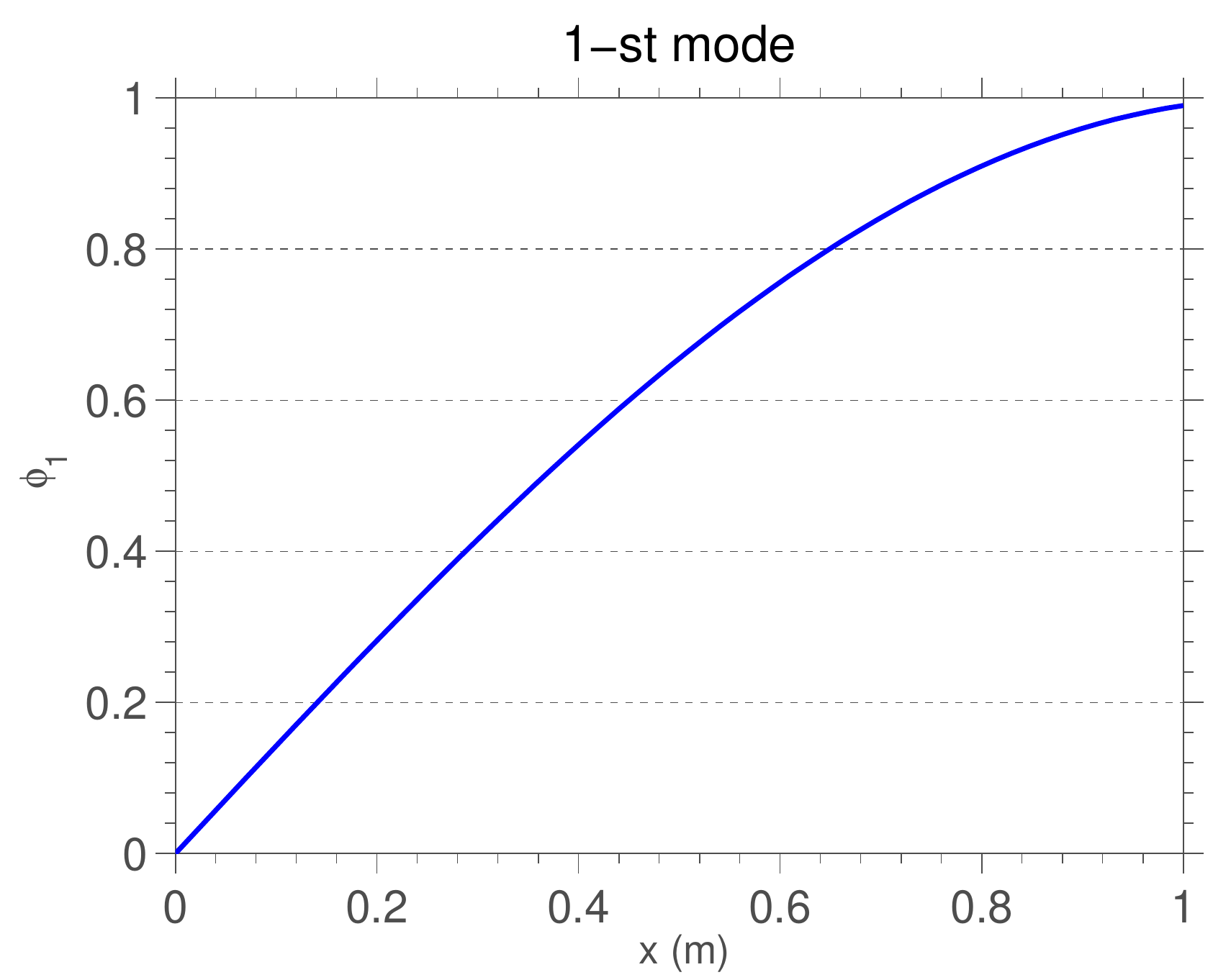}}
				\subfigure[nat. freq. $\approx 3.47~kHz$]{
				\includegraphics[scale=0.28]{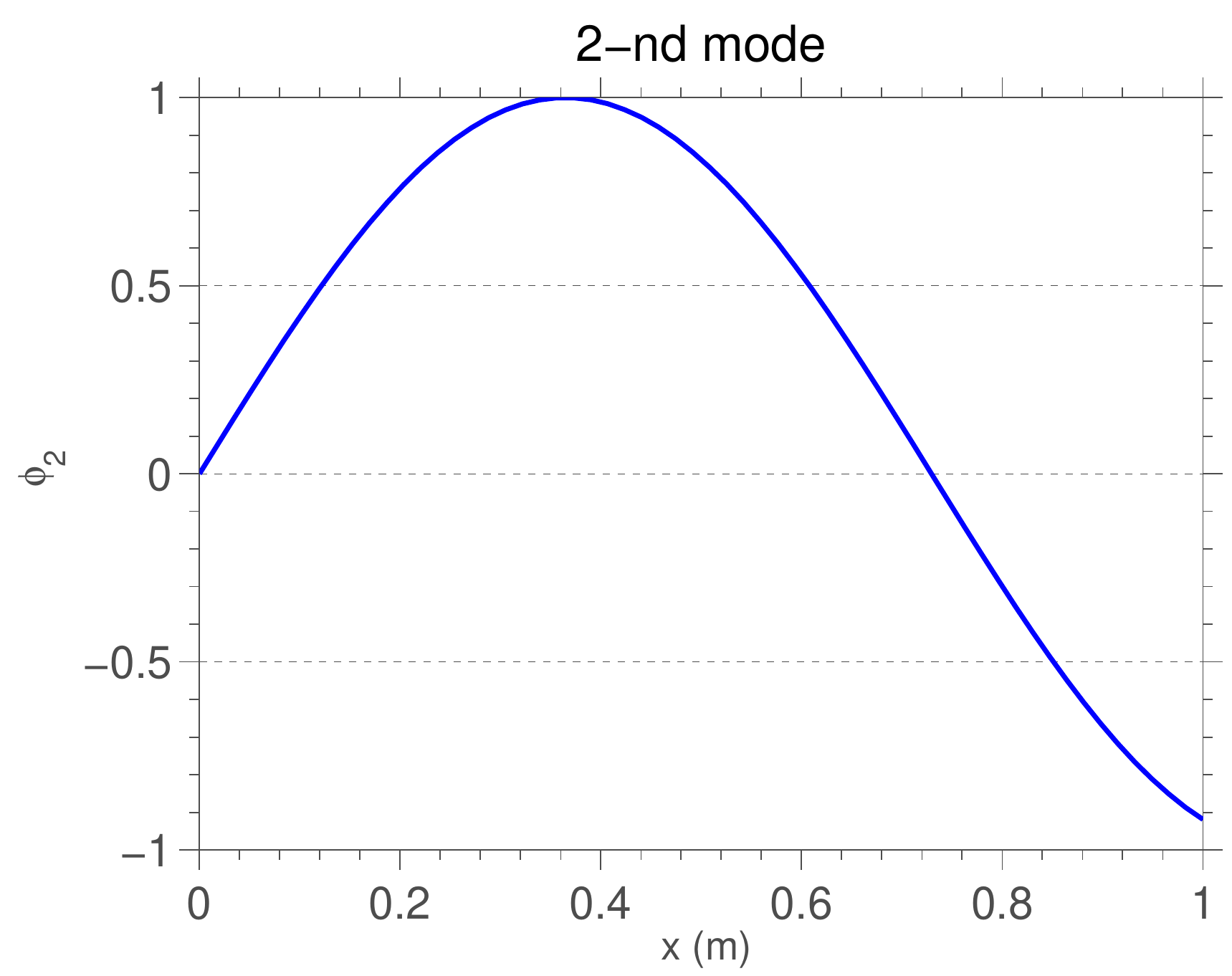}}\\
				\subfigure[nat. freq. $\approx 5.83~kHz$]{
				\includegraphics[scale=0.28]{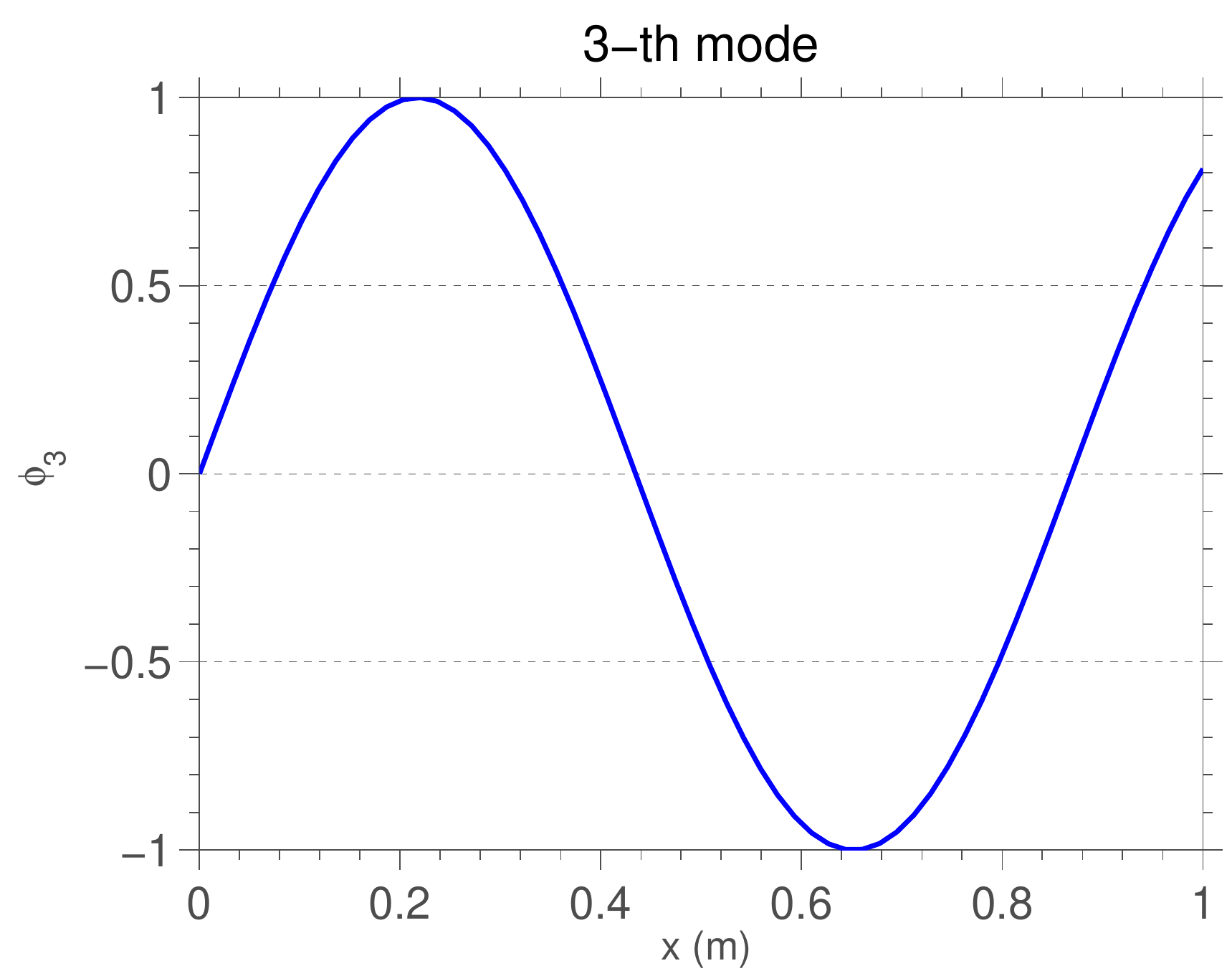}}
				\subfigure[nat. freq. $\approx 8.22~kHz$]{
				\includegraphics[scale=0.28]{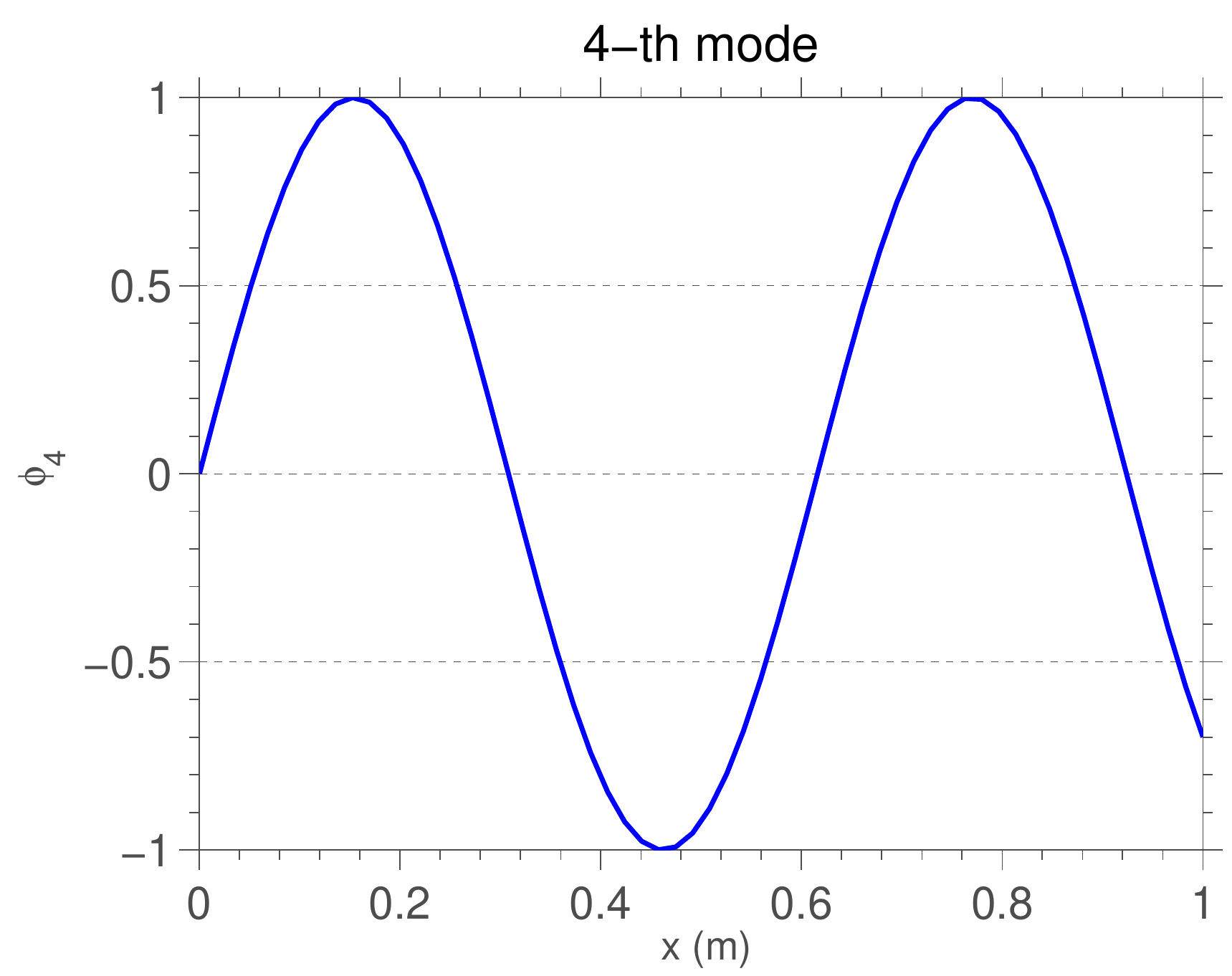}}\\
				\subfigure[nat. freq. $\approx 10.66~kHz$]{
				\includegraphics[scale=0.28]{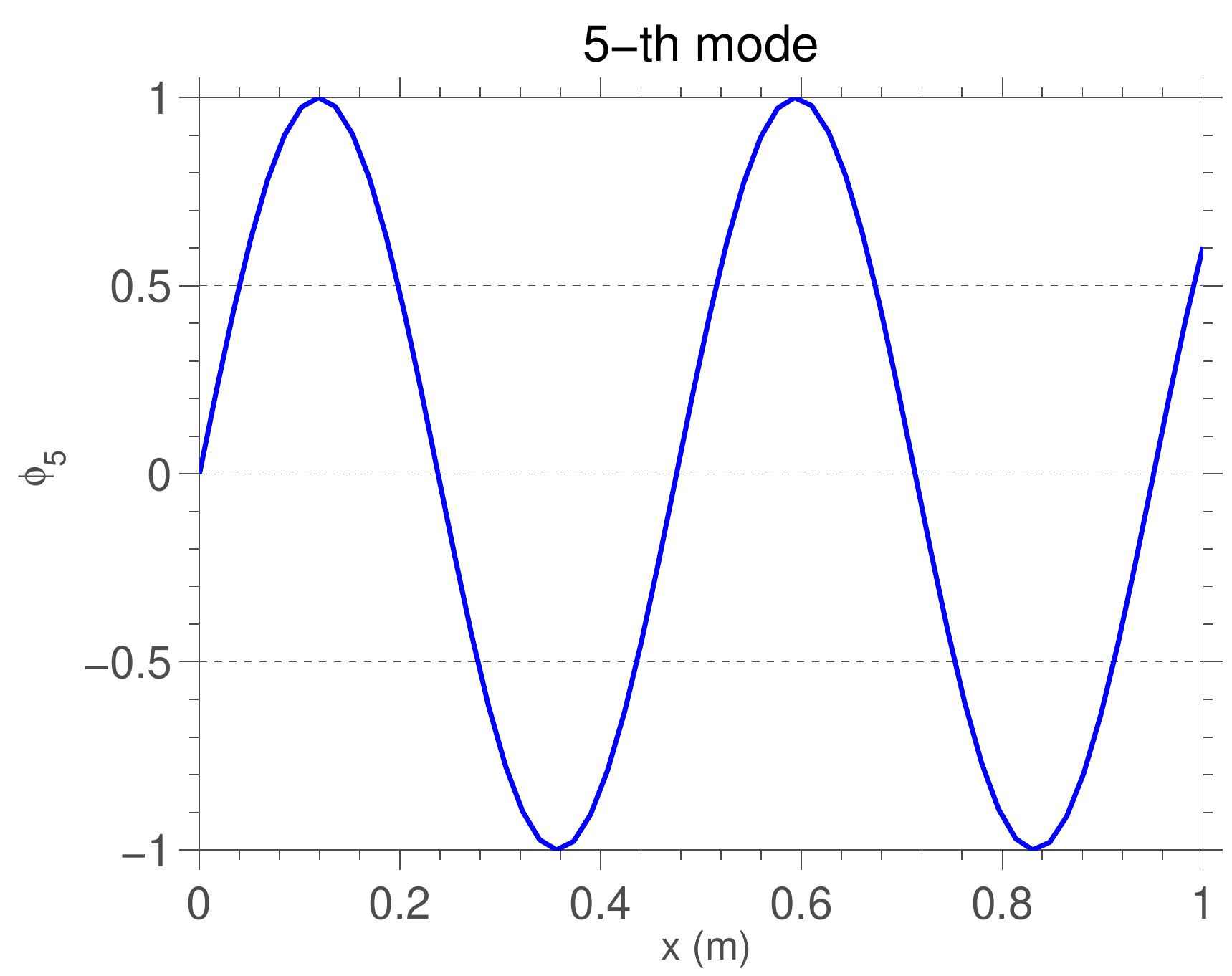}}
				\subfigure[nat. freq. $\approx 13.11~kHz$]{
				\includegraphics[scale=0.28]{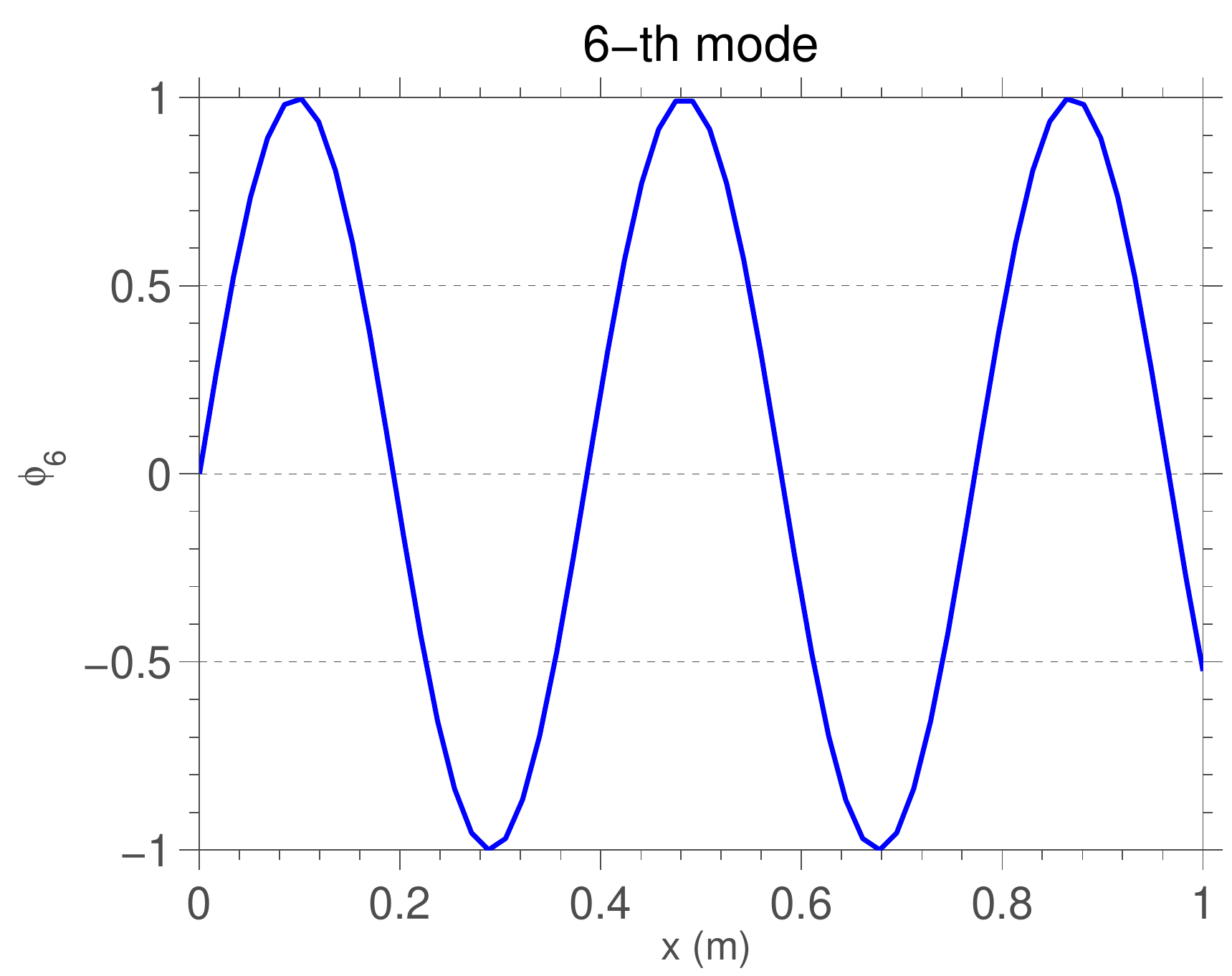}}
				\caption{The first six orthogonal modes and the corresponding 
				(approximated) natural frequencies of a fixed-mass-spring bar with $m=1.5~kg$.}
				\label{shape_modes_fig}
\end{figure}

The first six orthogonal modes shape of the fixed-mass-spring bar with $m=1.5~kg$,
whose the other parameters are presented in the beginning of section \ref{num_experim},
are illustrated in Figure~\ref{shape_modes_fig}. In this figure each sub-caption 
indicates the approximated natural frequency associated with the 
corresponding mode.

\subsection{Discretization of the model equations}
\label{galerkin_form}

The Galerkin method \citep{hughes2000} is employed to 
approximates the solution of the variational problem given by 
Eqs.(\ref{weak_eq}) to (\ref{weak_ic_eq2}). In this weighted
residual procedure the displacement field $u$ is approximated as

\begin{equation}
    u(x,t) \approx \sum_{n=1}^N u_n(t) \phi_n(x),
    \label{uN_eq}
\end{equation}

\noindent
where the basis functions $\phi_n$ are the orthogonal modes
of the conservative and non-forced dynamical system associated
to the fixed-mass-spring bar, and the coefficients $u_n$ 
are time-dependent functions. This results in the following
system of nonlinear ordinary differential equations

\begin{equation}
    \mtx{M} \ddot{\vec{u}}(t) + \mtx{C} \dot{\vec{u}}(t) +
    \mtx{K}\vec{u}(t) =  \vec{f}(t) + \vec{f}_{NL} \left( \dot{\vec{u}}(t) \right),
    \label{galerkin_eq}
\end{equation}

\noindent
supplemented by the following pair of initial conditions

\begin{equation}
    \vec{u}(t_0) =  \vec{u}_0 
    \qquad \mbox{and} \qquad 
    \vec{\dot{u}}(t_0) =  \vec{v}_0,
    \label{galerkin_ic_eq}
\end{equation}

\noindent
where $\vec{u}(t)$ is the vector of $\R^N$ in which the $n$-th component 
is $u_n(t)$, $\mtx{M}$ is the mass matrix, $\mtx{C}$ is the damping matrix, 
$\mtx{K}$ is the stiffness matrix.  Also, $\vec{f}(t)$, $\vec{f}_{NL}\left(\vec{u}(t)\right)$, 
$\vec{u}_0$, and $\vec{v}_0$ are vectors of $\R^N$, which respectively 
represent the external force, the nonlinear force, the initial position, 
and the initial velocity. The initial value problem of Eqs.(\ref{galerkin_eq}) and 
(\ref{galerkin_ic_eq}) has its solution approximated by Newmark method 
\citep{hughes2000}, in which a Newton-Raphson iteration is used to solve the 
nonlinear system of algebraic equations that arises from the discretization.


\section{Stochastic modeling of the mechanical system}
\label{stochastic_approach}

\subsection{Stochastic initial--boundary value problem}

Consider a probability space  $(\SS, \SF, \PM)$, 
where $\SS$ is sample space, 
$\SF$ is a $\sigma$-field over $\SS$ and 
$\PM$ is a probability measure. 
In this probability space, the elastic modulus 
is assumed to be a random variable $\randvar{E}: \SS \to \R$,
and the distributed external force a random field 
$\randproc{F}: [0,L] \times [t_0,t_f]  \times \SS \to \R$.

Due to the randomness of $\randproc{F}$ and 
$\randvar{E}$, the bar displacement becomes a random field
$\randproc{U}: [0,L] \times [t_0,t_f]  \times \SS \to \R$, 
which evolves according to

\begin{eqnarray}
    \rho A \dpd[2]{\randproc{U}}{t} +  c \dpd{\randproc{U}}{t} & = &
    \dpd{}{x} \left( \randvar{E} A \dpd{\randproc{U}}{x} \right) \\ \nonumber
    & - & \left( k \randproc{U} + k_{NL} \randproc{U}^3 + m \dpd[2]{\randproc{U}}{t} \right) \delta(x - L)
    + \randproc{F} (x,t,\SSpt).
    \label{randbar_eq}
\end{eqnarray}

This problem has boundary and initial conditions similar to those 
defined in Eqs.(\ref{bar_bc})~and~(\ref{bar_ic}), by changing 
$u$ for $\randproc{U}$ only. Furthermore, the partial derivatives 
now are not defined in the classical way, but in the mean square
sense \citep{papoulis2002}.

\subsection{Random external force modeling}

The distributed external force acting on the bar is assumed
as the form

\begin{equation}
	\randproc{F}(x,t,\SSpt) = \sigma \phi_1(x) N(t,\SSpt),
	\label{ext_force}
\end{equation}

\noindent
where $\sigma$ is the force amplitude, $\phi_1$ the bar first mode\footnote{The choice 
of the spatial shape of the excitation seek for a configuration that is physically plausible and simple. 
The first mode meets both requirements.},
and $N(t,\SSpt)$ is a Gaussian white-noise\footnote{Remember that a white-noise 
is a random process which all instants of time are uncorrelated. In other words, 
the behavior of the process at any given instant of time has no influence on the other instants.}
with zero mean and unit variance. 

\begin{figure}[h!]
		\centering
		\includegraphics[scale=0.31]{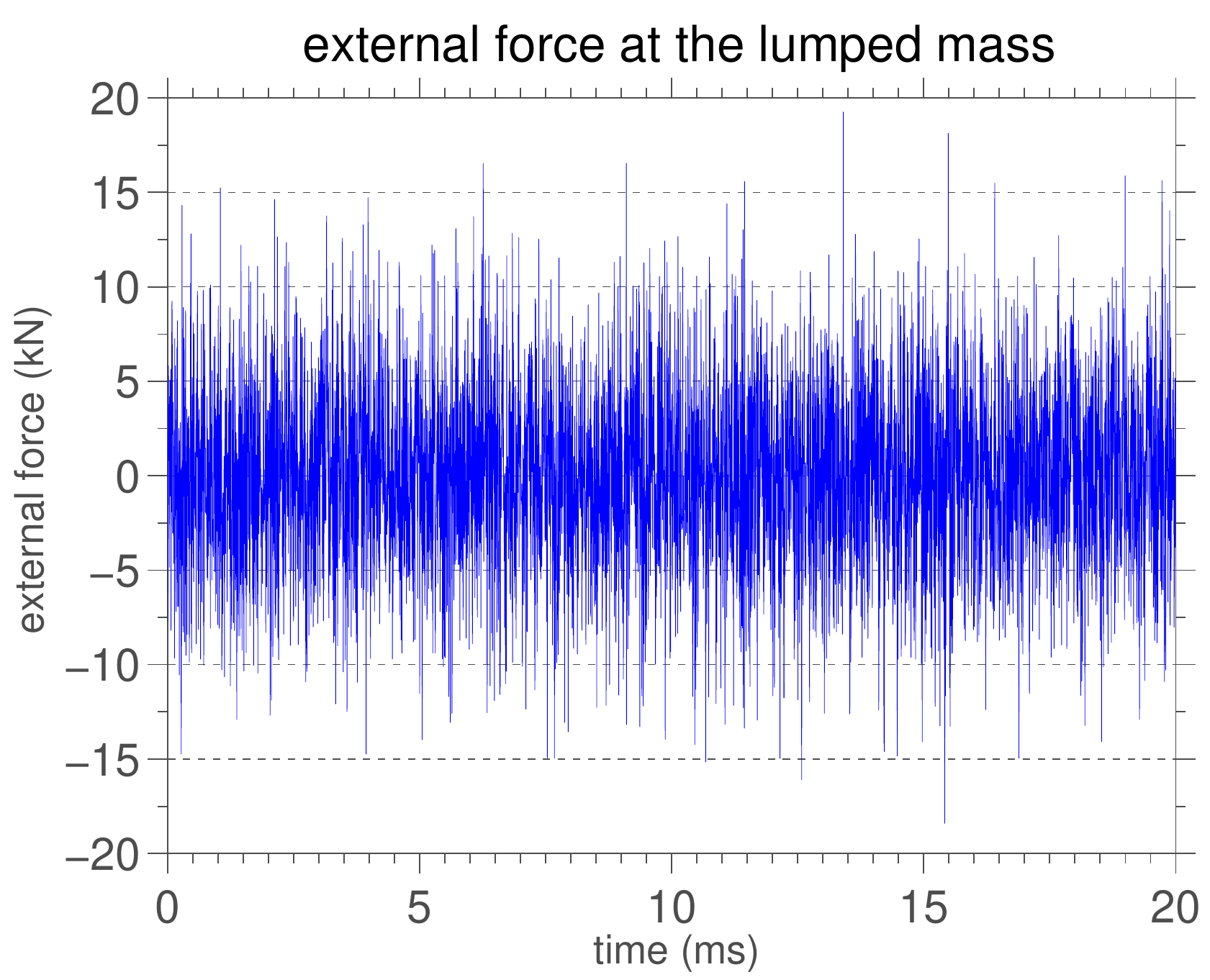}
		\caption{This figure illustrates a realization of the random external force at $x=L$.}
		\label{ext_force_fig}
\end{figure}

A typical realization of the random external force, given by Eq.(\ref{ext_force}), 
for fixed position, is shown in Figure~\ref{ext_force_fig}.

\subsection{Random elastic modulus distribution}

The elastic modulus cannot be negative, so it is reasonable to assume 
the support of $\randvar{E}$ as the interval $(0,\infty)$. 
Therefore, the probability density function (PDF) of $\randvar{E}$ 
is a nonnegative function $\pdf{\randvar{E}}: (0,\infty) \to \R$, which 
respects the following normalization condition

\begin{equation}
	\int_{0}^{\infty} \pdf{\randvar{E}}(\xi) \, d\xi = 1.
	\label{norm_cond_E}
\end{equation}

Additionally, it is supposed that the expected value of $\randvar{E}$ 
is a known (finite) real number, i.e.,

\begin{equation}
	\expvalop{\randvar{E}}{0}{\infty} = \mu_{\randvar{E}} < \infty,
	\label{meanval_E_finite}
\end{equation}

\noindent
as well as the expected value of  $\ln{\left( \randvar{E} \right)}$,

\begin{equation}
	\int_{0}^{\infty} \ln{\left( \randvar{E} \right)} \, \pdf{\randvar{E}}(\xi) \, d\xi
	= \mu_{\randvar{\ln{\left( \randvar{E} \right)}}} < \infty,
	\label{meanval_log_E_finite}
\end{equation}

\noindent
being the latter requirement a sufficient condition to ensure that 
$\randvar{E^{-1}}$ exists almost sure, and is a second order random variable 
\citep{soize2000p277, soize2005p1333}.

Following the suggestion of \cite{soize2000p277, soize2005p1333,soize2013p2379},
the maximum entropy principle is employed in order 
to consistently specify $\pdf{\randvar{E}}$.
This methodology chooses for $\randvar{E}$ the PDF
which maximizes the entropy function defined by

\begin{equation}
	\entropy{\pdf{\randvar{E}}} = \entropyop{\pdf{\randvar{E}}}{0}{\infty},
	\label{dif_entropy}
\end{equation}

\noindent
subjected to the constraints given by (\ref{norm_cond_E}), (\ref{meanval_E_finite})
and (\ref{meanval_log_E_finite}).
These restrictions effectively define the known information about $\randvar{E}$.

The gamma distribution is the one which solves the optimization
problem above, and its PDF is given by

\begin{equation}
	\pdf{\randvar{E}}(\xi) = \indfunc{(0,\infty)}
	\frac{1}{\mu_{\randvar{E}}} 
	\left( \frac{1}{\delta_{\randvar{E}}^2} \right)^{ \left( \displaystyle \frac{1}{\delta_{\randvar{E}}^2} \right) }
	\frac{1}{\Gamma(1/\delta_{\randvar{E}}^2)} 
	\left( \frac{\xi}{\mu_{\randvar{E}}} \right)^{ \left( \displaystyle \frac{1}{\delta_{\randvar{E}}^2}-1 \right) }
	\exp\left( - \frac{\xi}{\delta_{\randvar{E}}^2 \mu_{\randvar{E}}} \right),
	\label{gamma_distrib}
\end{equation}

\noindent
where $\indfunc{(0,\infty)}$ denotes the indicator function of the 
interval $(0,\infty)$, $\Gamma$ indicates the gamma function, and
$\delta_{\randvar{E}}$ is a type of dispersion parameter, such that
$1 \leq \delta_{\randvar{E}} \leq 1/\sqrt{2}$, defined as the ratio 
between the standard deviation and the mean of $\randvar{E}$.

\subsection{Stochastic solver: Monte Carlo method}

Uncertainty propagation in the nonlinear stochastic dynamics of the bar
is computed by Monte Carlo (MC) method \cite{casella2010,cunhajr2014p1355}. 
This stochastic solver uses a pseudorandom number generator
to obtain many realizations of $\randvar{E}$ and $\randproc{F}$. 
Each one of these realizations defines a new Eq.(\ref{weak_eq}), 
so that a new weak problem is obtained. After that, these new weak 
problems are solved deterministically, such as in section~\ref{galerkin_form}.
All the MC simulations reported in this work use $4096$ samples
to access the random system.


\section{Numerical experimentation}
\label{num_experim}

The numerical experiments presented in this section adopt for the system 
parameters the deterministic values shown in Table~\ref{param_table}.
Also, the random variable $\randvar{E}$, is characterized by the mean
$\mu_{\randvar{E}} = 203~GPa$ and the dispersion $\delta_{\randvar{E}} = 0.1$.

\begin{table}[h!]
\caption{This table presents the deterministic (nominal) parameters 
used in the numerical simulations reported in this work.}
\vspace{2mm}
\label{param_table}
\centering
	\begin{tabular}{cll}
		\toprule
		parameter & value & unit  \\
		\cmidrule(r){1-3}
		$\rho$ & $7900$ & $kg/m^3$ \\
		$A$ & $625\pi$ & $mm^2$ \\
		$L$ & $1$ & $m$ \\
		$c$ & $5$ & $kN/s$ \\
		$k$ & $650$ & $N/m$ \\
		$k_{NL}$ & $650 \times 10^{13}$ & $N/m^3$ \\
		$\sigma$ & $5$ & $kN$ \\
		$\alpha_1$ & $0.1$ & $mm$ \\
		$\alpha_2$ & $0.5 \times 10^{-3}$ & --- \\
		\bottomrule
	\end{tabular}
\end{table}

The approximation to the solution of the weak initial-boundary value problem
of section~\ref{weak_form}, constructed as described in the section~\ref{galerkin_form},
uses 10 modes. As the 10-th natural frequency of the system is $\approx 23.08~kHz$, 
a representative frequency band of this dynamical system is 
$\mathfrak{B} = [0,25]~kHz$. Thus, to analyze the dynamics of the system in this 
frequency band, it is adopted a ``temporal window" given by the interval 
$[t_0,t_f] = [0,20]~ms$.

For sake of reference, a deterministic (nominal) model, with $E=\mu_{\randvar{E}}$,
and $f(x,t) = \sigma \phi_1(x)$, is considered. Furthermore, a parametric study, 
with  $\dimless{m} = 0.1,~1,~10,~50$, 
is performed to investigate the effect of the end mass on the bar dynamics, where
the discrete--continuous mass ratio is defined as

\begin{equation}
		\dimless{m} = \frac{m}{\rho A L}.
\end{equation}

\subsection{Evolution of the lumped mass velocity}
\label{numexp__velo}

\begin{figure} [h!]
				\centering
				\subfigure[$\dimless{m} = 0.1$]{
				\includegraphics[scale=0.31]{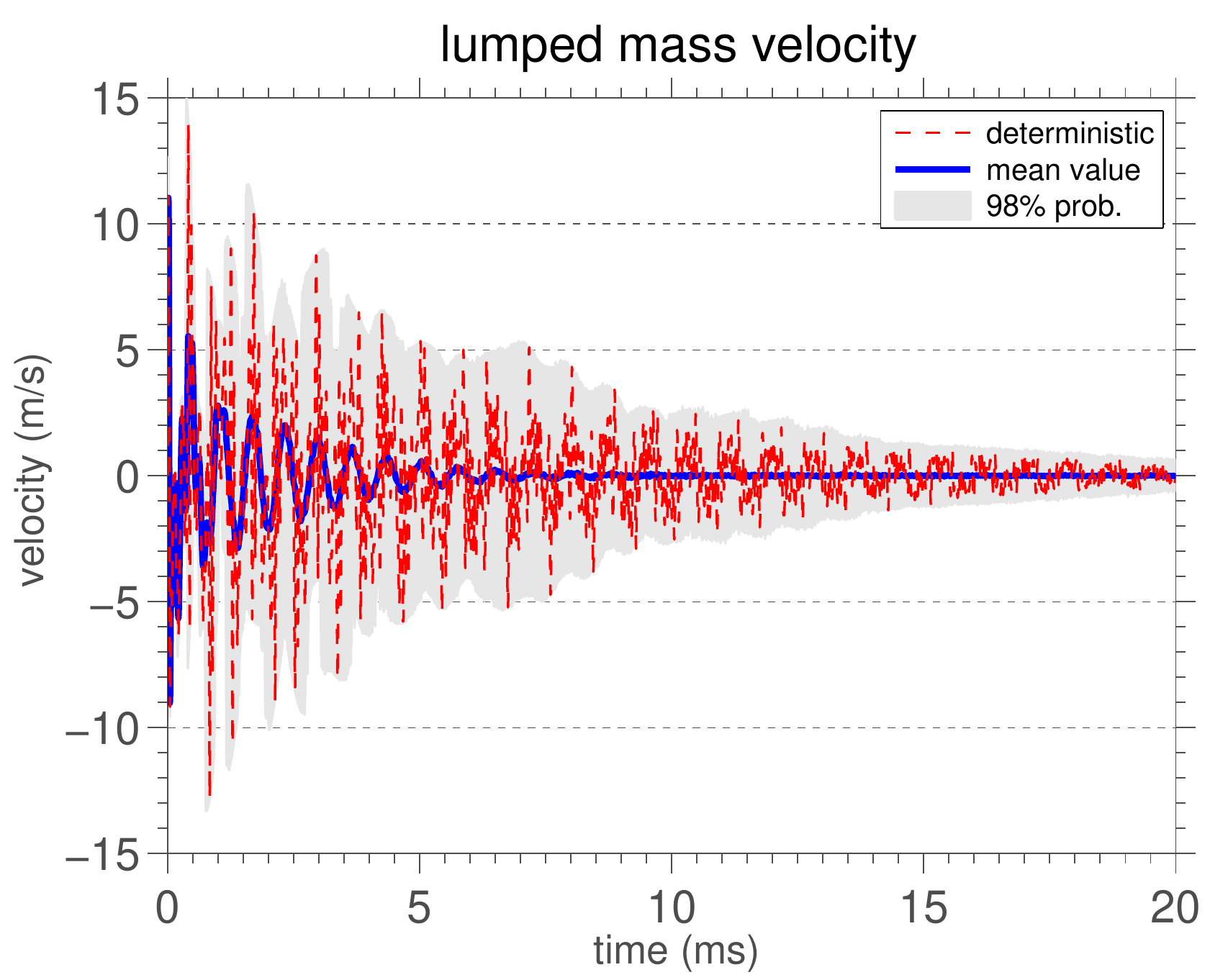}}
				\subfigure[$\dimless{m} = 1$]{
				\includegraphics[scale=0.31]{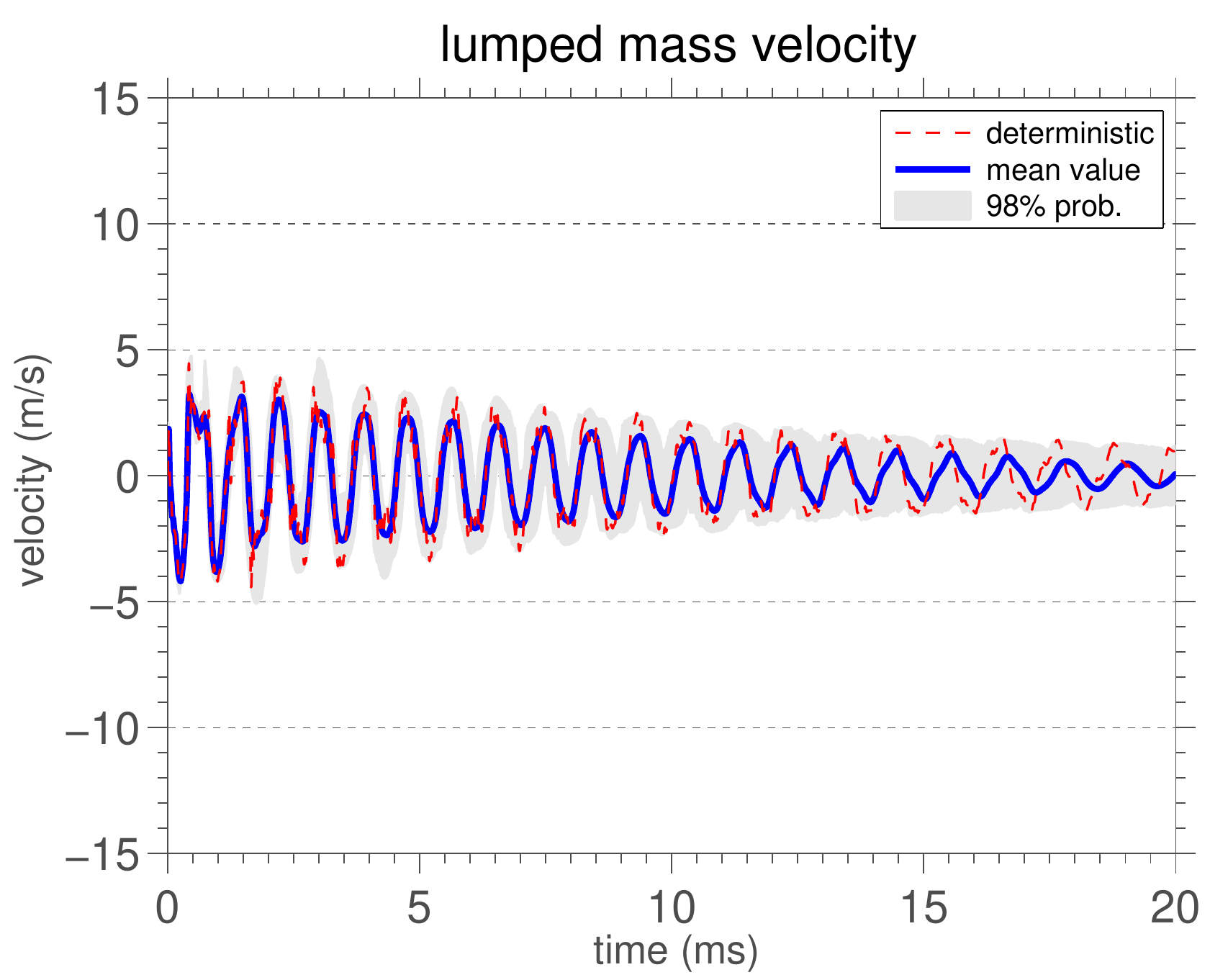}}\\
				\subfigure[$\dimless{m} = 10$]{
				\includegraphics[scale=0.31]{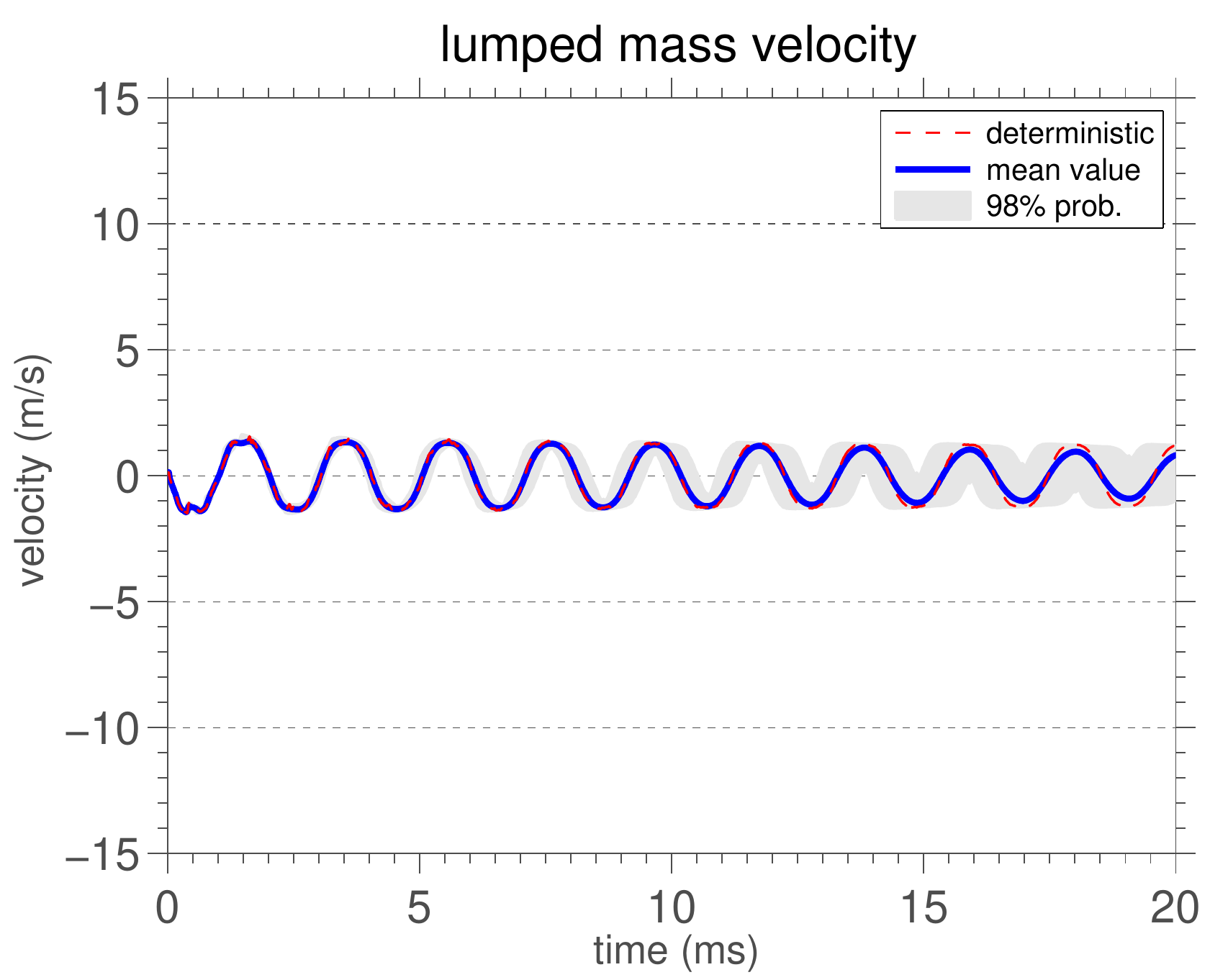}}
				\subfigure[$\dimless{m} = 50$]{
				\includegraphics[scale=0.31]{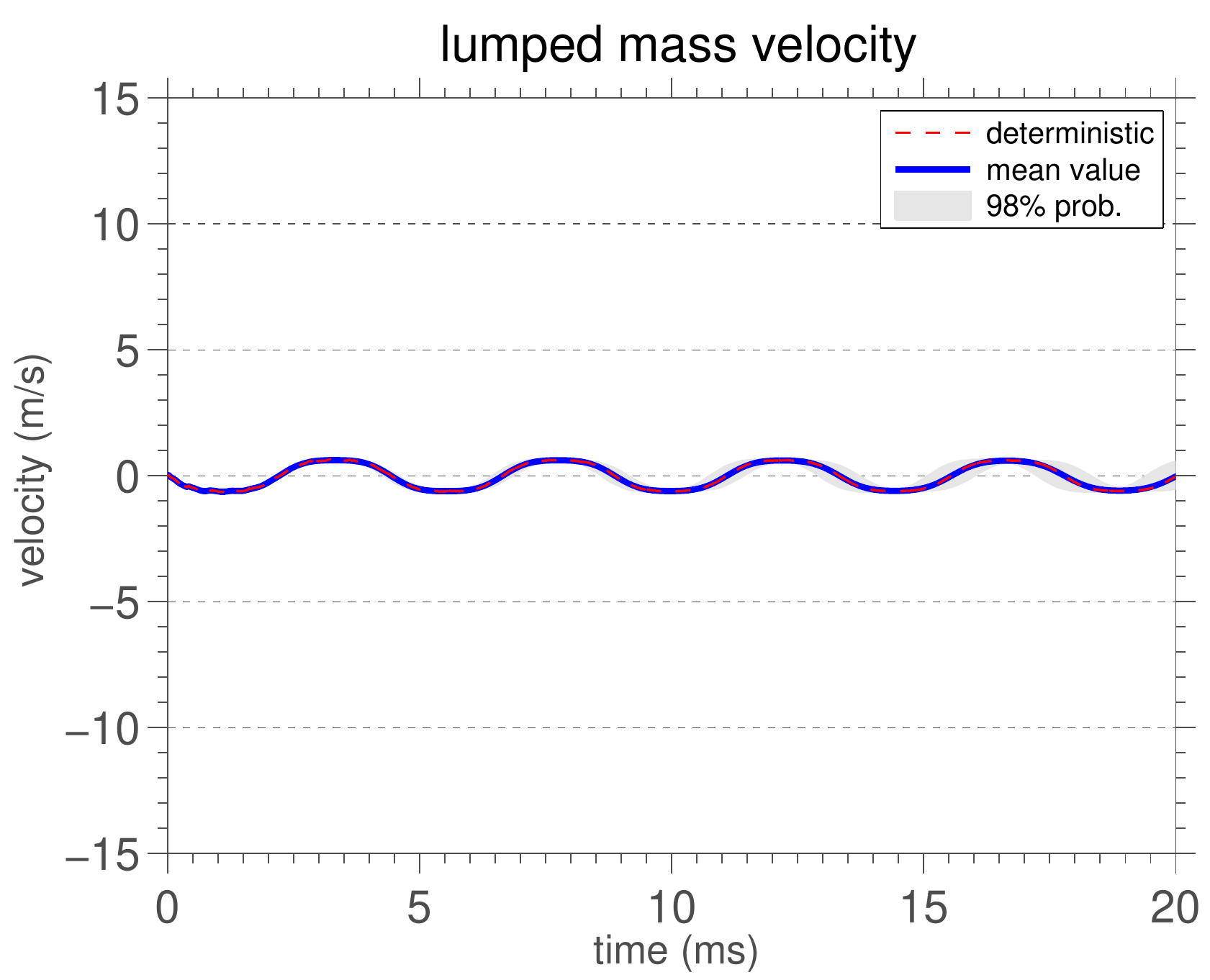}}
				\caption{This figure illustrates the mean value (blue line) and a 98\% of probability interval 
							  of confidence (grey~shadow) for the random process $\dot{\randproc{U}}(L,\cdot,\cdot)$,
							  for several values of the discrete--continuous mass ratio.}
				\label{ci_utL_fig}
\end{figure}

The mean value of the lumped mass velocity, i.e, $\dot{\randproc{U}}(L,\cdot,\cdot)$, 
its nominal value, and an envelope of reliability, wherein a realization of the stochastic 
system has 98\% of probability of being contained, are shown, for different values of 
$\dimless{m}$, in Figure~\ref{ci_utL_fig}. By observing this figure one can note that, 
as the value of lumped mass increases, the mean value tends to the nominal value. 
That is, the system is ``more random" for small values of $\dimless{m}$. 

Also, the analysis of Figure~\ref{ci_utL_fig} shows that, for large values of $\dimless{m}$, 
the decay in the system displacement amplitude decreases significantly, i.e., the system 
is not much influenced by damping as $\dimless{m} \to \infty$.

Explanations for the observations made in the preceding paragraphs of this section 
are provided by the analysis of the system orbit in phase space, which is done in
the section~\ref{numexp__phase_space}.

Furthermore, the amplitude of the confidence interval increases with time 
for all values of $\dimless{m}$, i.e., the system uncertainty at $x=L$
is greater in the stationary regime. This is evident in the first three 
graphs, but remains true in the fourth graph, and is due to the accumulation 
of uncertainties with the increasing time.

\subsection{Orbit of the lumped mass in the mechanical system phase space}
\label{numexp__phase_space}

\begin{figure} [ht!]
				\centering
				\subfigure[$\dimless{m} = 0.1$]{
				\includegraphics[scale=0.31]{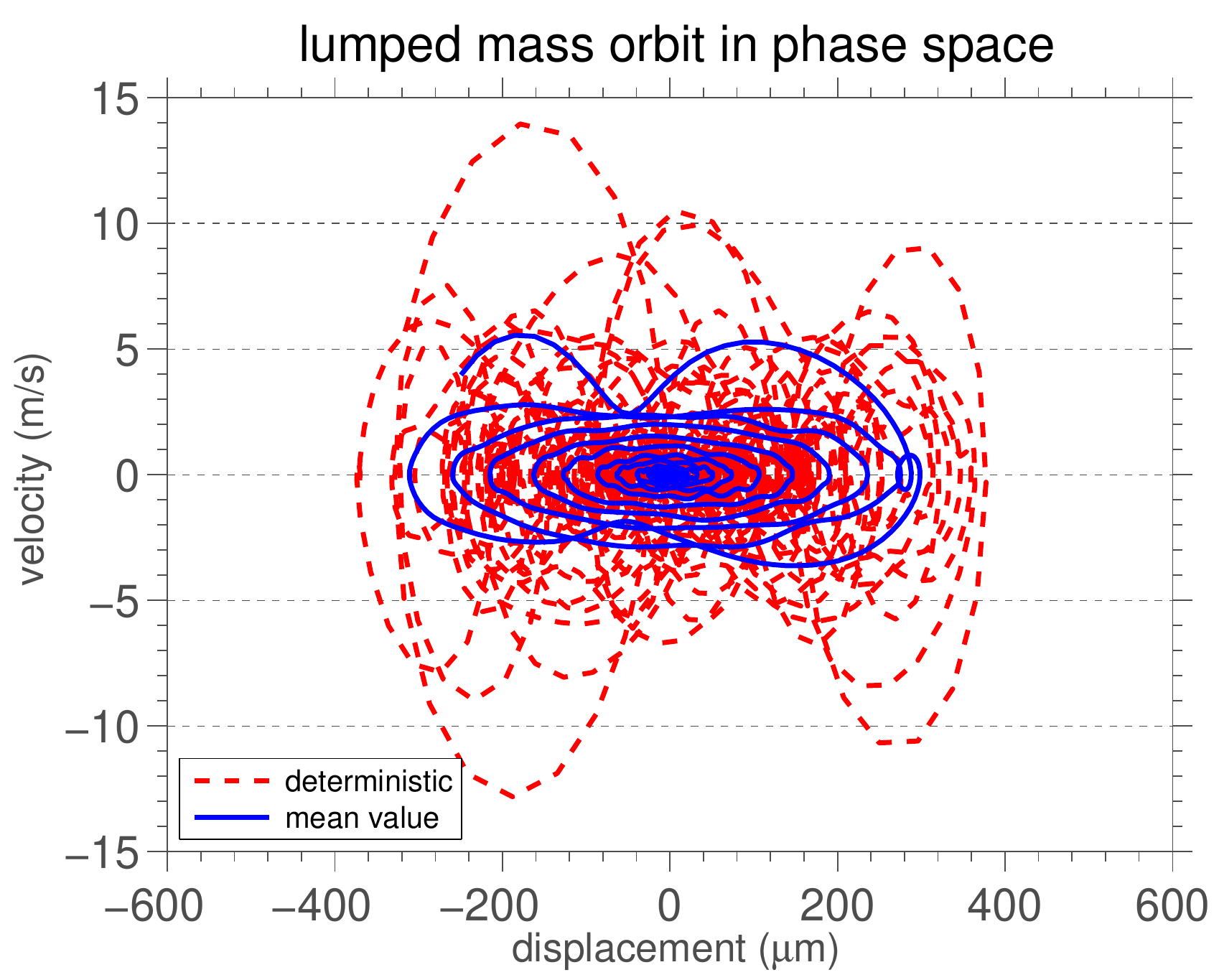}}
				\subfigure[$\dimless{m} = 1$]{
				\includegraphics[scale=0.31]{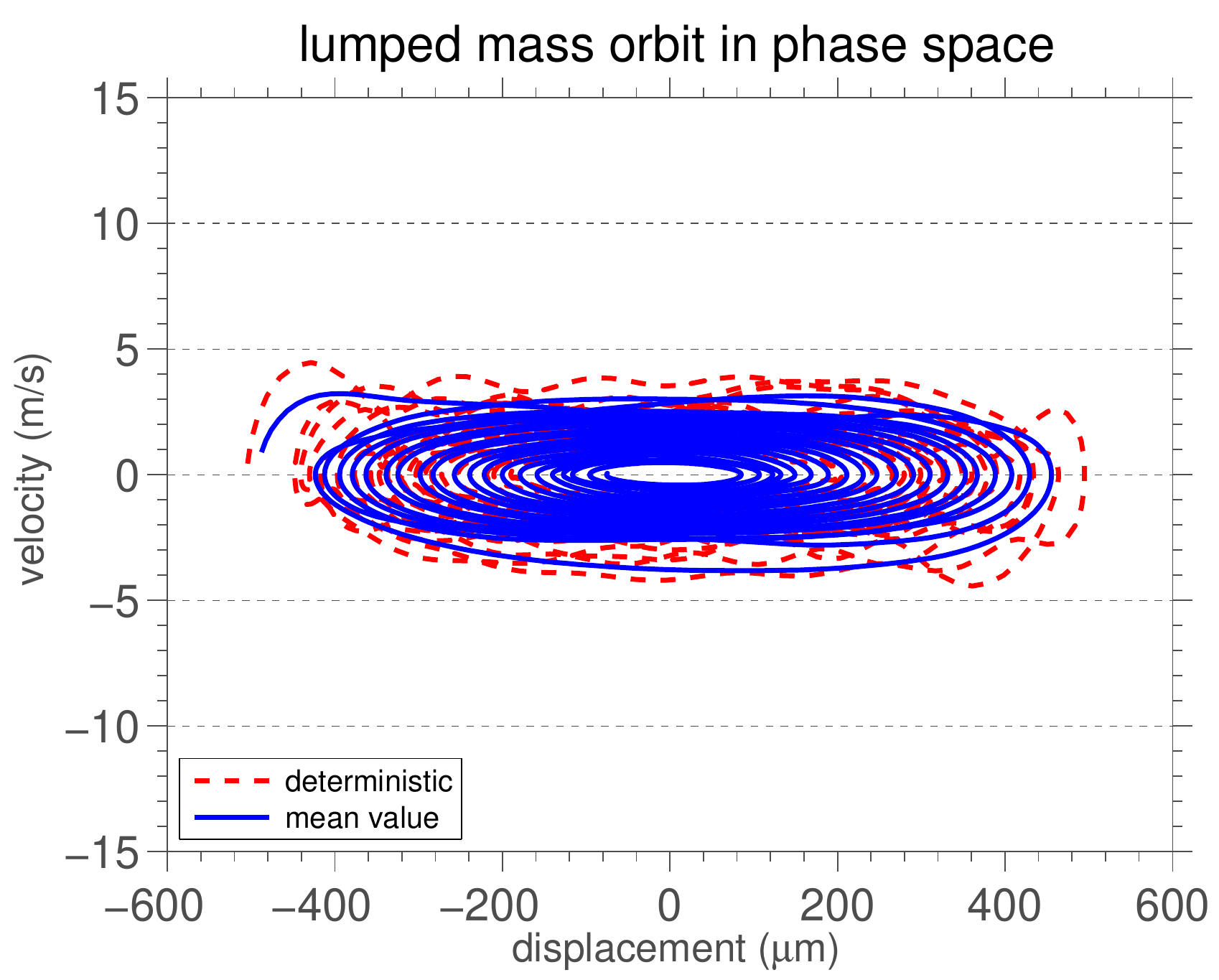}}\\
				\subfigure[$\dimless{m} = 10$]{
				\includegraphics[scale=0.31]{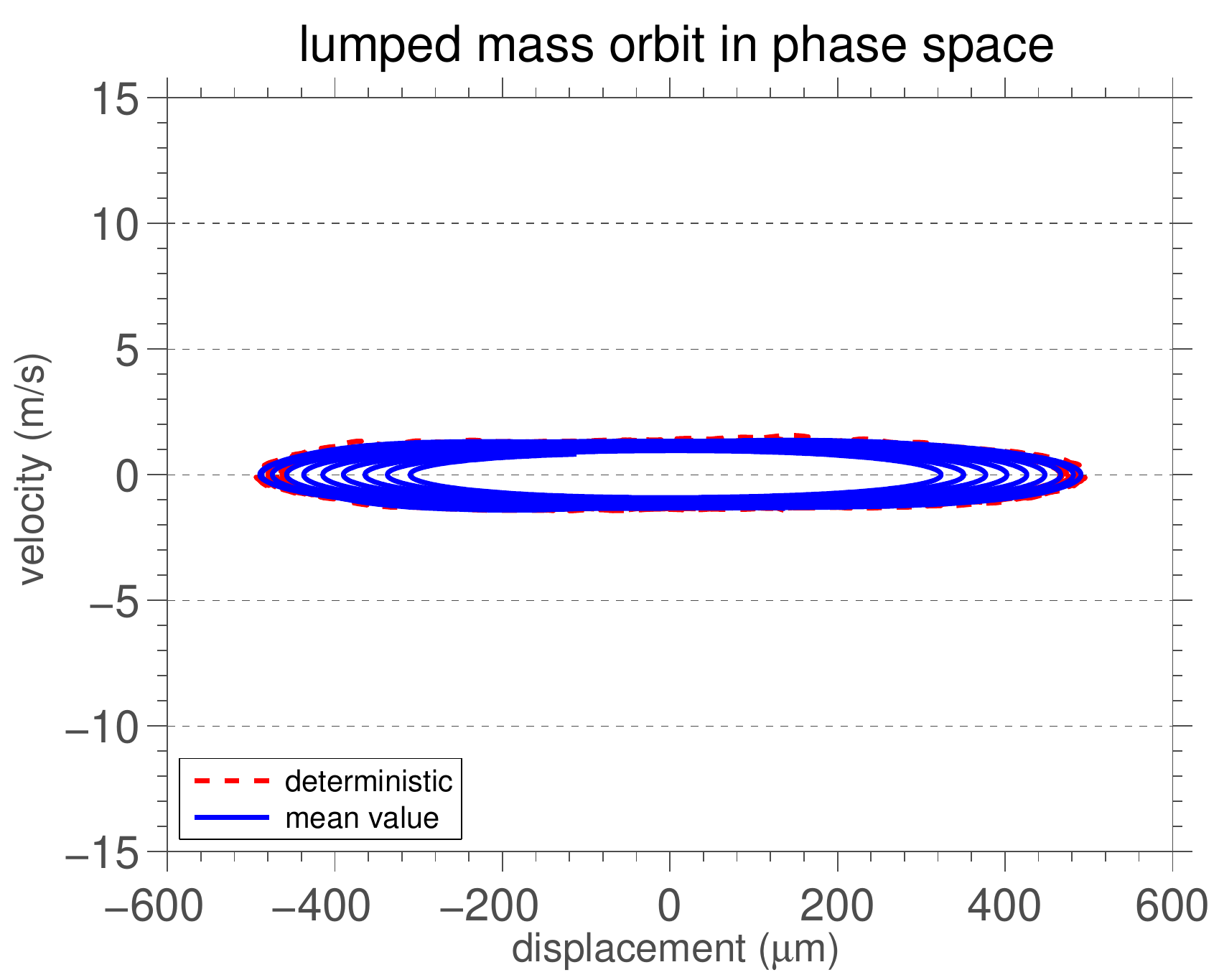}}
				\subfigure[$\dimless{m} = 50$]{
				\includegraphics[scale=0.31]{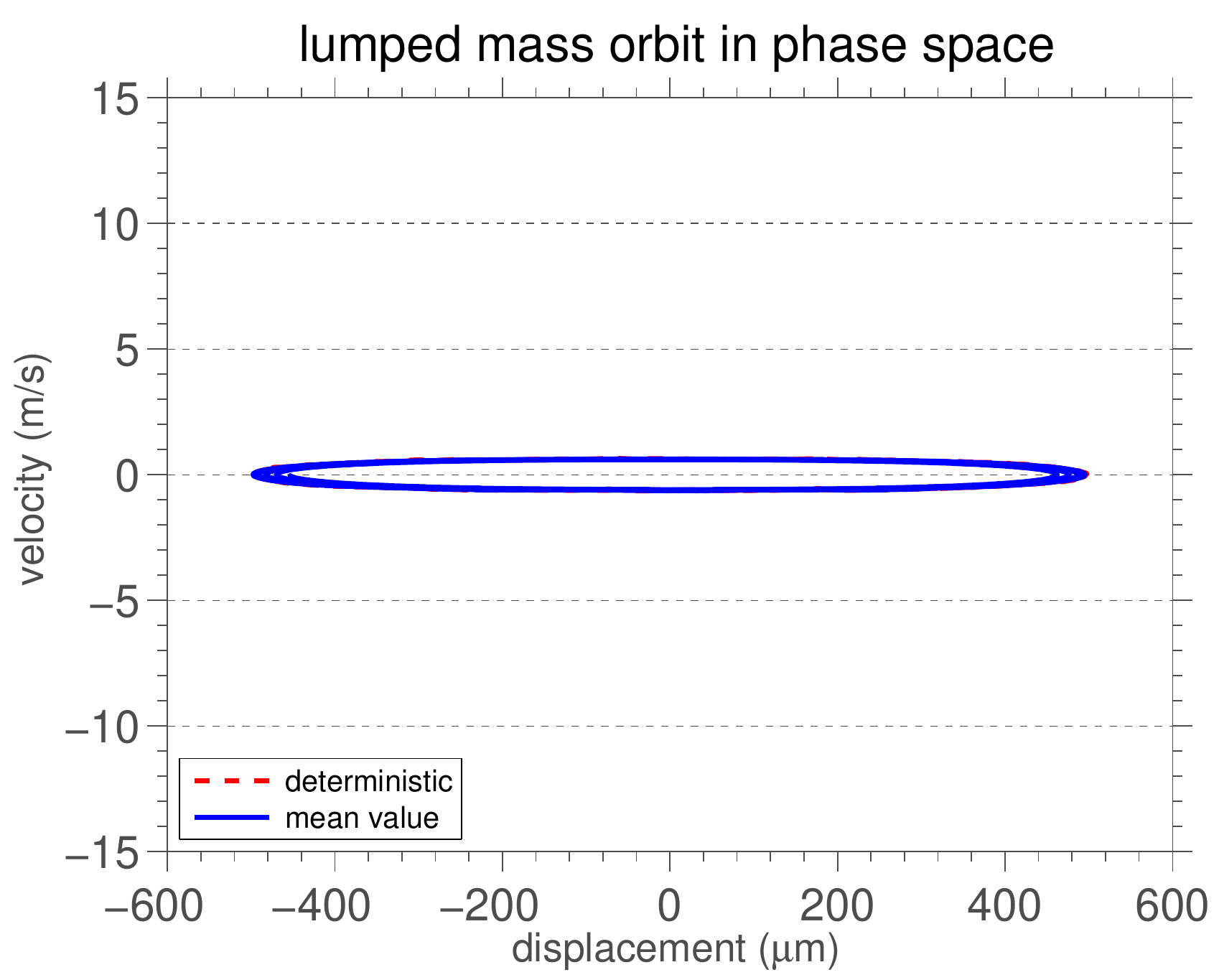}}
				\caption{This figure illustrates the mean orbit, in the phase space, of the 
							  fixed-mass-spring bar at $x=L$, for several values of the 
							  discrete--continuous mass ratio.}
				\label{phase_space_fig}
\end{figure}

The mean orbit, in the phase space, of the fixed-mass-spring bar at $x=L$ is shown, 
for different values of $\dimless{m}$, in Figure~\ref{phase_space_fig}. Distinct behaviors,
for the different values of $\dimless{m}$ shown, can be observed.

For $\dimless{m}=0.1$, the mean orbit is quite different from the 
``disturbed" nominal orbit observed. This is because the response of the nominal
system depends on the initial conditions for a long period, fact which is not observed
for the other values of $\dimless{m}$. This explains why the mean velocity
tends to the nominal velocity when $\dimless{m}$ increases.

The assertive made in the second paragraph of section~\ref{numexp__velo},
about how the influence of the damping in the system decreases, can be confirmed 
by analyzing the Figure~\ref{phase_space_fig}, since the mean orbit of 
the system tends from a stable focus to an ellipse as $\dimless{m}$ increases. 
So, the limit behavior of the bar right extreme with $\dimless{m} \to \infty$ 
is a mass-spring system. This limit behavior, which tends to a conservative system, 
occurs because, with the increasing of $\dimless{m}$, most of the mass of the system 
becomes concentrated at the right extreme of the bar. Thus, the bar behaves like a massless 
spring. Also, as the damping is distributed along the bar and the mass of it became negligible, 
the viscous dissipation becomes ineffective.

\subsection{Power spectral density of the lumped mass velocity}

\begin{figure} [ht!]
				\centering
				\subfigure[$\dimless{m} = 0.1$]{
				\includegraphics[scale=0.31]{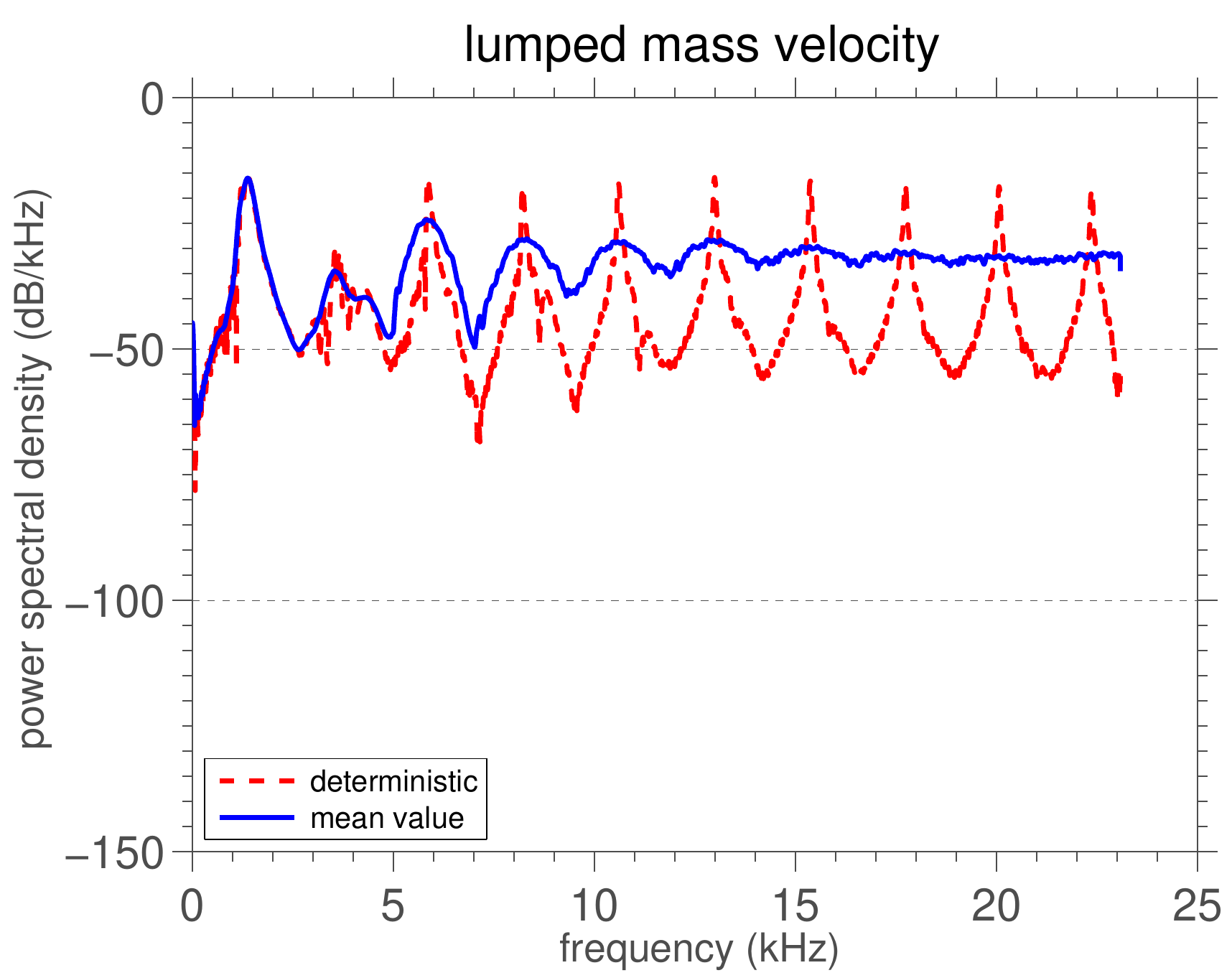}}
				\subfigure[$\dimless{m} = 1$]{
				\includegraphics[scale=0.31]{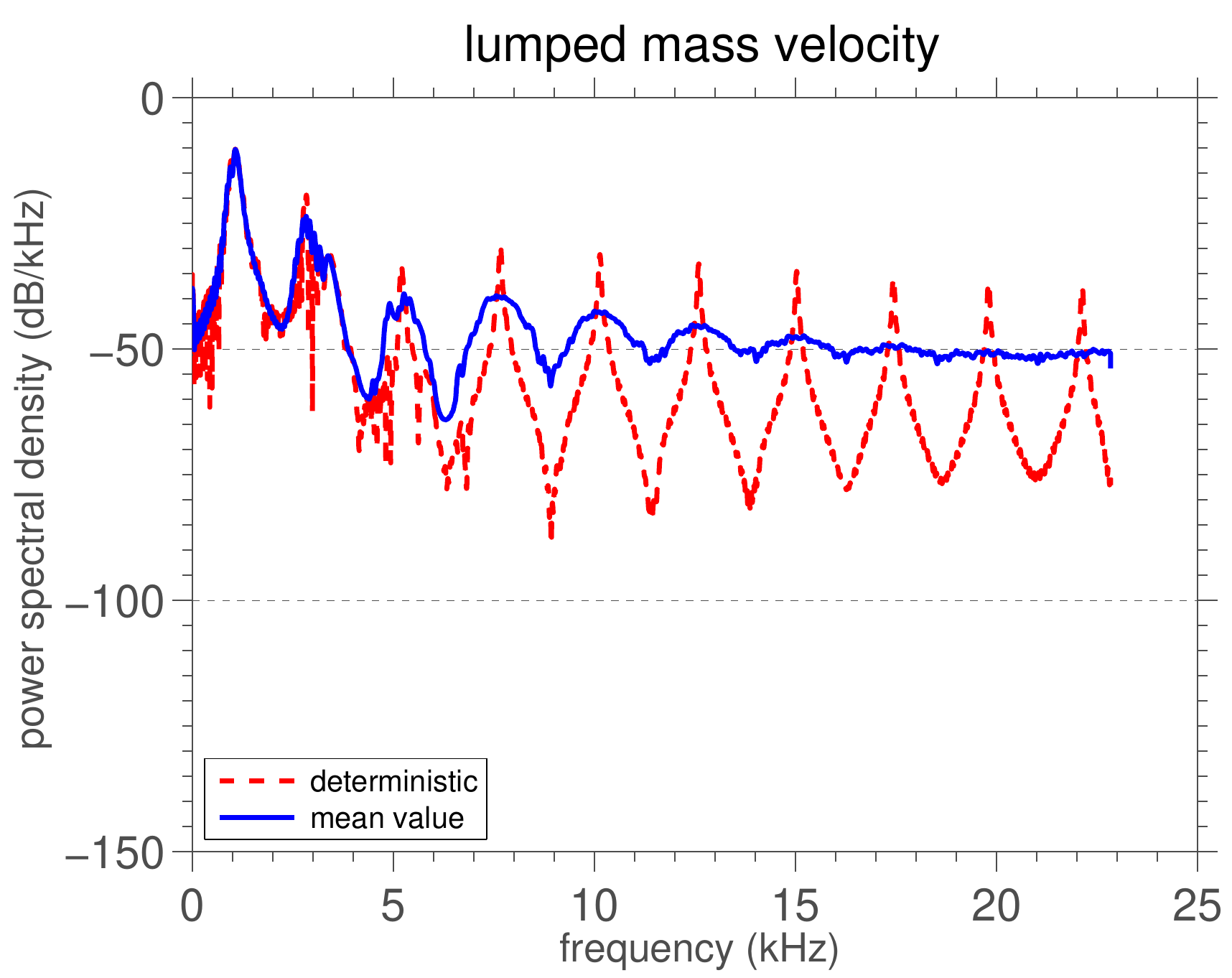}}\\
				\subfigure[$\dimless{m} = 10$]{
				\includegraphics[scale=0.31]{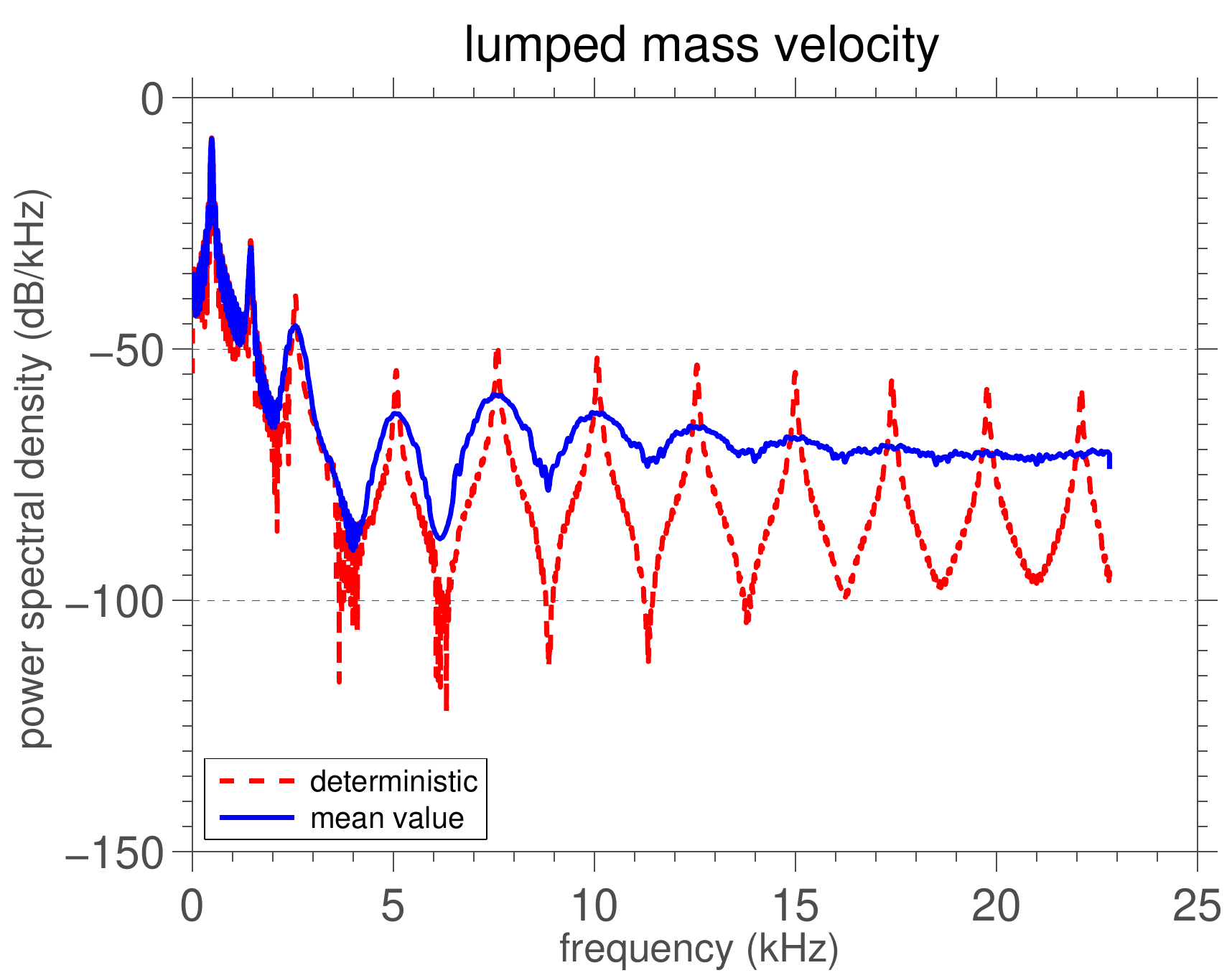}}
				\subfigure[$\dimless{m} = 50$]{
				\includegraphics[scale=0.31]{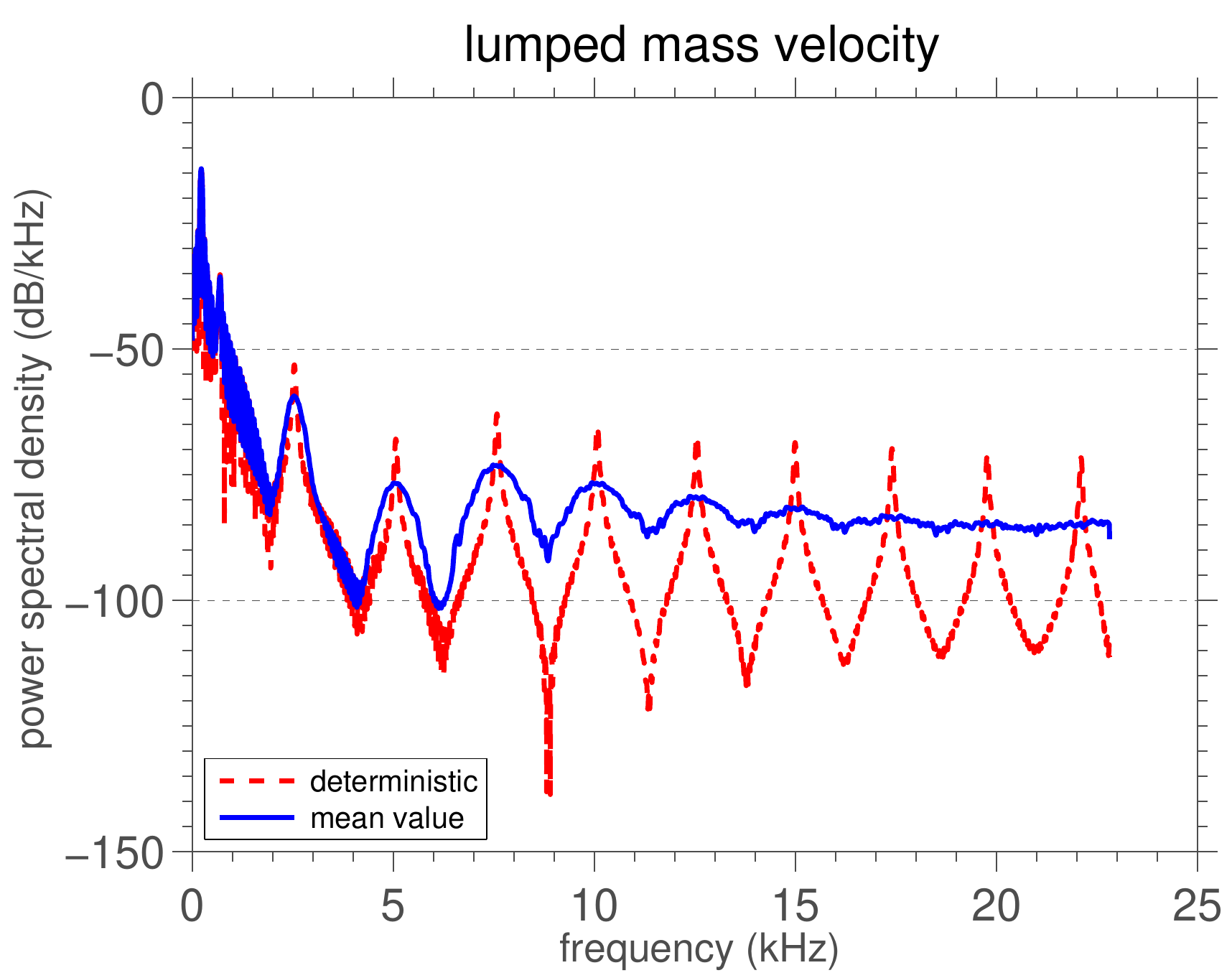}}
				\caption{This figure illustrates estimations to the PSD of the random process 
				$\dot{\randproc{U}}(L,\cdot,\cdot)$, for several values of the discrete--continuous mass ratio.}
				\label{spectral_density_fig}
\end{figure}

The energy distribution of the bar through the frequency spectrum can be 
seen in Figure~\ref{spectral_density_fig}, which shows the mean power 
spectral density (PSD) of the lumped mass (steady state) velocity and 
its nominal value.

The presence of the white-noise forcing excites the mechanical system 
in all frequencies of the band $\mathfrak{B}$. This is made evident by the 
various peaks in the mean PSD function, each one occurring in a frequency 
that is very close to a natural frequency of the system.
It is important to note that the peaks of the nominal and of the mean
PSD occur practically at the same frequencies. Once the forcing does 
not influence the natural frequencies, the only random parameter to 
promote changes in natural frequencies is $\randvar{E}$, whose the 
randomness is reasonably low.

\pagebreak
A larger number of peaks can be seen in the high frequencies, 
but the peak with greater height, and thus, the more energy, is always the
first frequency of the spectrum. As the spatial dependence of the forcing 
is given by the first mode, as can be seen in Eq.(\ref{ext_force}), the low 
frequency of the spectrum receives an ``extra contribution" of energy beyond 
the white-noise.

However, as $\dimless{m}$ increases, the natural frequencies of the associated
conservative system decrease, which is not observed in the case of the bar.
This difference in the system behavior, as well as irregular redistribution
of energy along the spectrum, when $\dimless{m}$ changes, may be due to 
cubic nonlinearity.

\subsection{Probability density function of the lumped mass velocity}
\label{numexp__pdf}

\begin{figure} [ht!]
				\centering
				\subfigure[$\dimless{m} = 0.1$]{
				\includegraphics[scale=0.31]{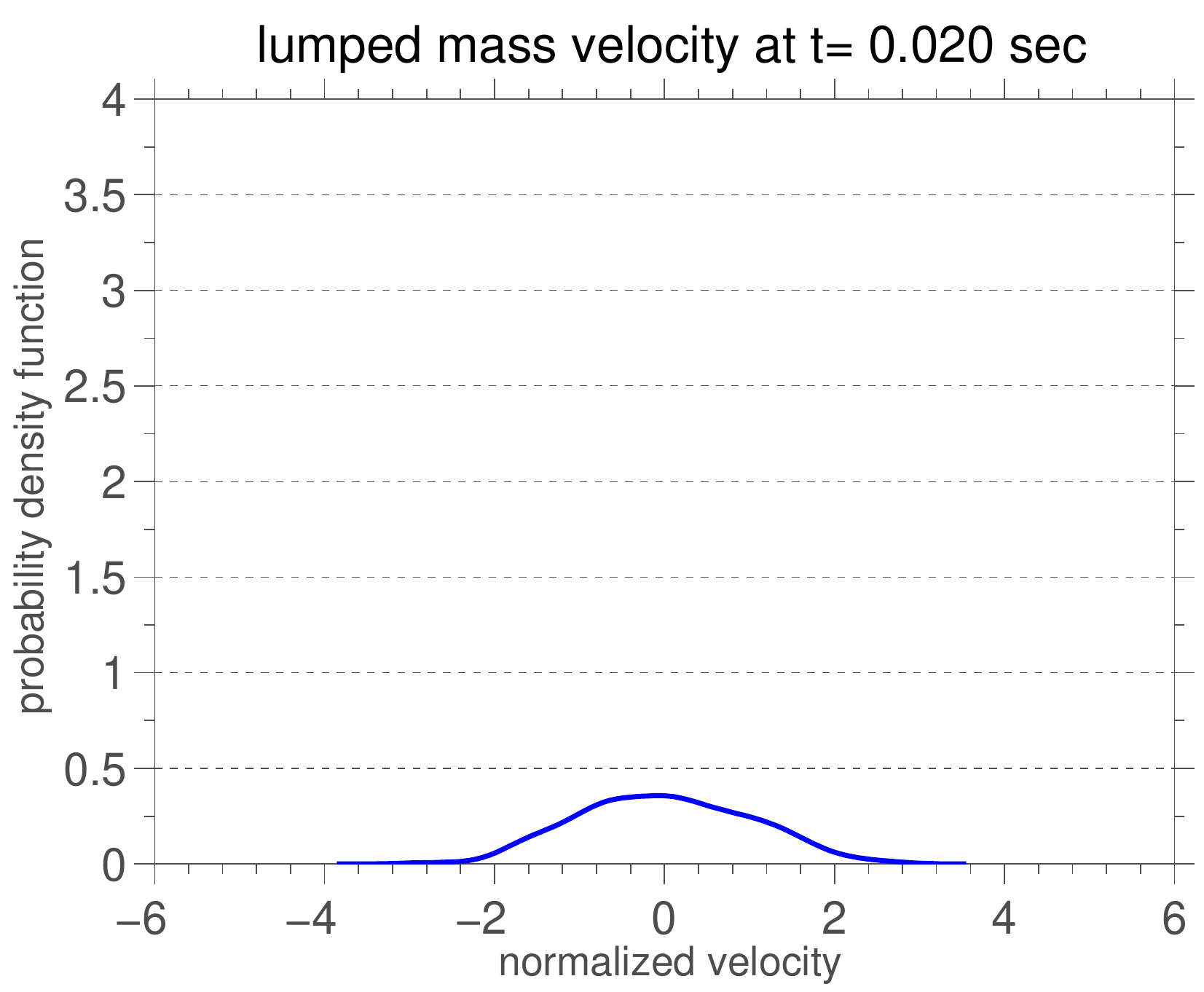}}
				\subfigure[$\dimless{m} = 1$]{
				\includegraphics[scale=0.31]{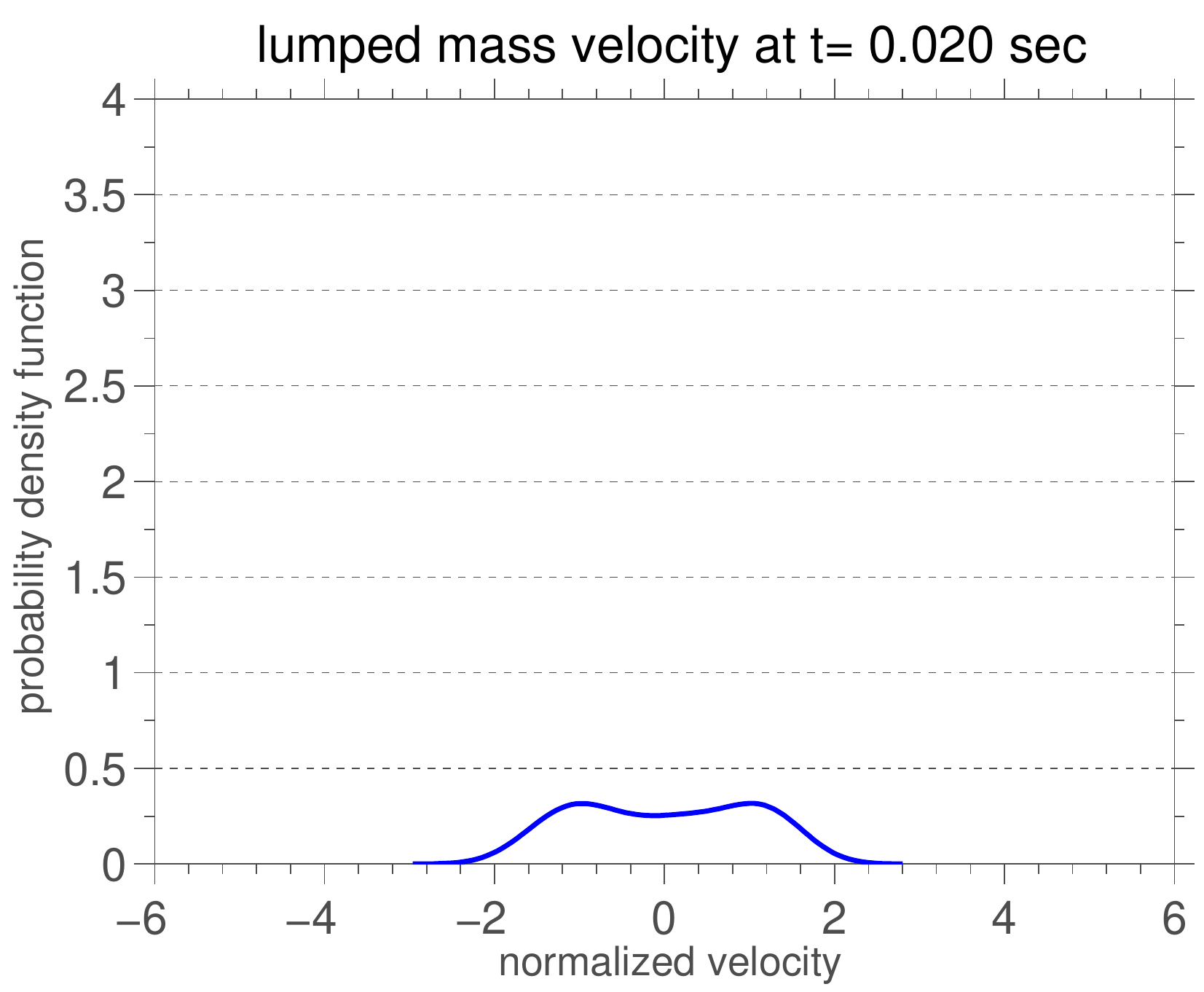}}\\
				\subfigure[$\dimless{m} = 10$]{
				\includegraphics[scale=0.31]{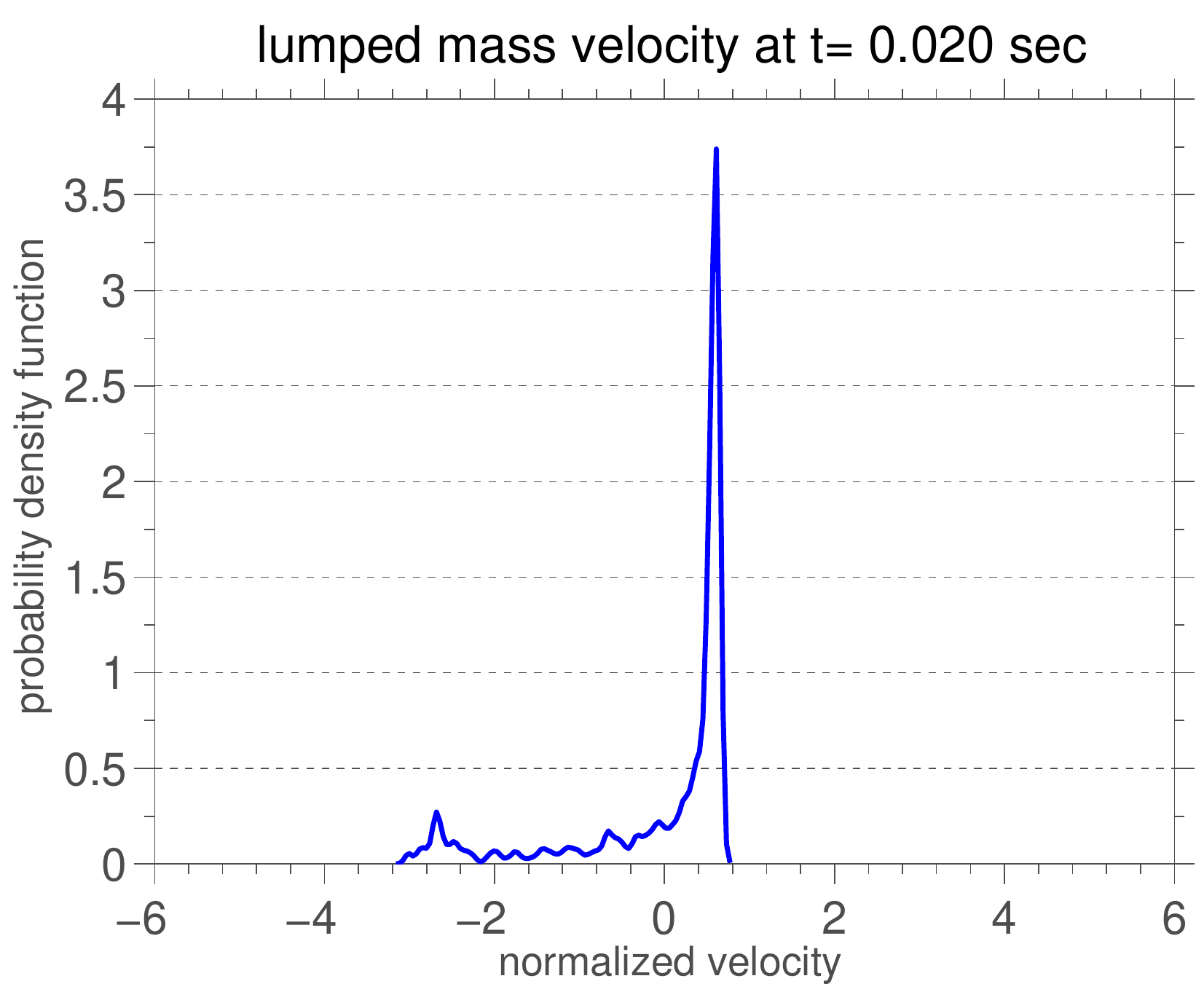}}
				\subfigure[$\dimless{m} = 50$]{
				\includegraphics[scale=0.31]{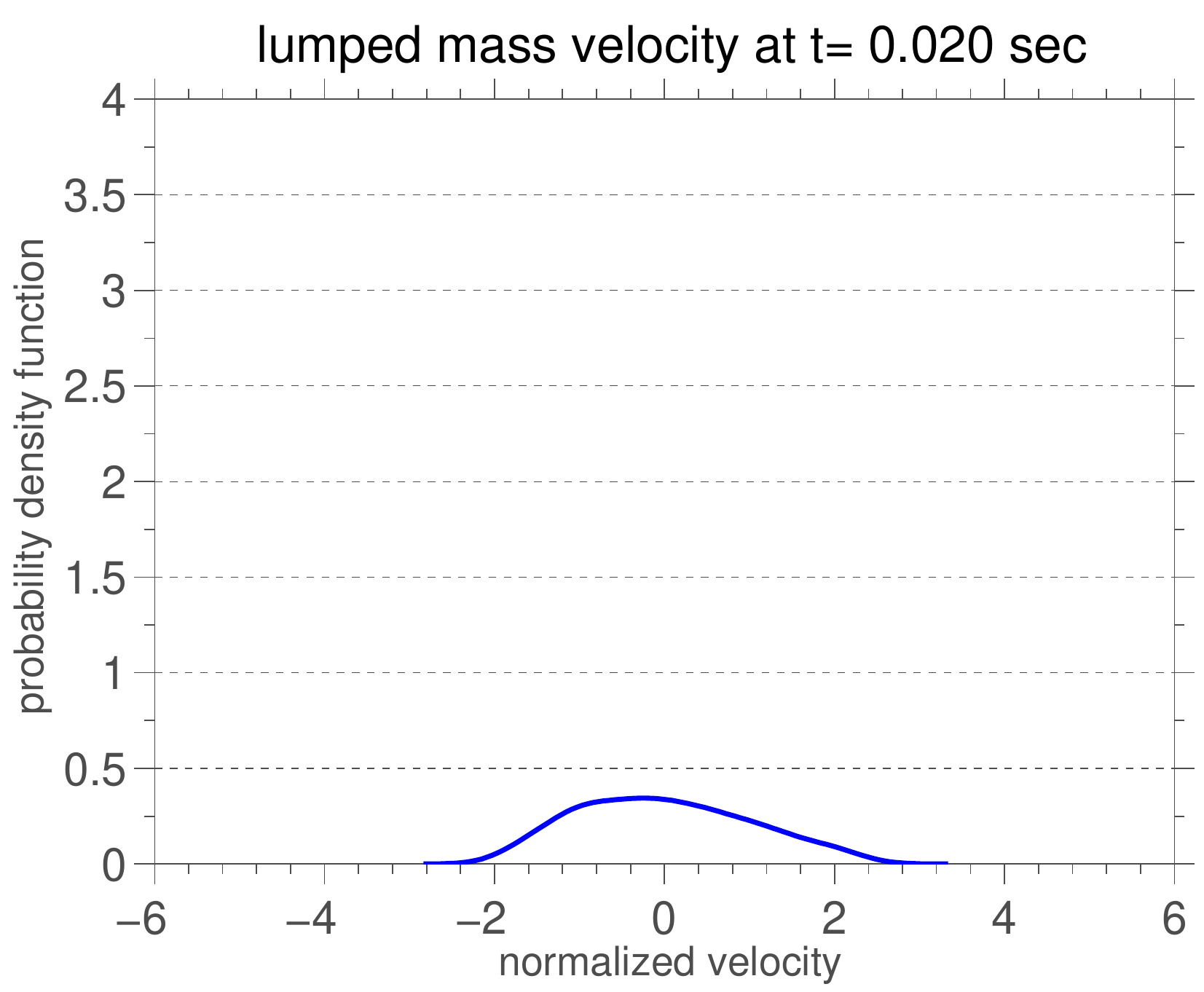}}
				\caption{This figure illustrates estimations to the PDF of the (normalized)
								random variable $\dot{\randproc{U}}(L,t_f,\cdot)$, for several values 
								of the discrete--continuous mass ratio.}
				\label{pdf_utL_fig}
\end{figure}

The difference between the system dynamical behavior is even clearer if one looks at
the PDF estimations\footnote{These estimates were obtained using a kernel smooth 
density technique \cite{bowman1997}.} of the normalized random variable 
$\dot{\randproc{U}}(L,t_f,\cdot)$, which are presented in Figure~\ref{pdf_utL_fig}.
Note that in this context normalized means a random variable with zero mean and 
unit standard deviation.

In all cases the PDF presents an asymmetry around the (zero) mean
as can be seen in the Table~\ref{asymmetry_table}, which shows the 
probability of the normalized random variable $\dot{\randproc{U}}(L,t_f,\cdot)$ 
be less than or equal to the mean, for several values of the discrete--continuous 
mass ratio.

\pagebreak
These asymmetries indicate if it is more probable the velocity be higher or 
lower than the mean, according to the area under the PDF curve to the right 
or to the left of the mean, respectively. The observed values are in agreement 
with what is seen in the envelopes of reliability of Figure~\ref{ci_utL_fig}.

\begin{table}[h!]
\caption{This table presents the probability of the normalized 
random variable $\dot{\randproc{U}}(L,t_f,\cdot)$ be less than or equal to the mean,
for several values of the discrete--continuous mass ratio.}
\vspace{2mm}
\label{asymmetry_table}
\centering
	\begin{tabular}{cc}
		\toprule
		$\dimless{m}$ &  probability\\
		\cmidrule(r){1-2}
		0.1 & $\thickapprox$ 0.52 \\
		1 & $\thickapprox$ 0.50 \\
		10 & $\thickapprox$ 0.28 \\
		50 & $\thickapprox$ 0.53 \\
		\bottomrule
	\end{tabular}
\end{table}

Furthermore, it is possible to observe a multimodal behavior in some of
the PDFs shown in the Figure~\ref{ci_utL_fig}.
This multimodal behavior indicates a high number 
of realizations close to the values that correspond to the peaks.
Therefore, it can be concluded that the regions near the peaks
are areas of greater probability for the system response.
Note that these areas change irregularly when $\dimless{m}$
is varied.


\section{Concluding remarks}
\label{concl_remaks}

This work presents a model to describe the nonlinear dynamics of a elastic bar,
attached to discrete elements, with viscous damping, random elastic modulus, 
and subjected to a Gaussian white-noise distributed external force.
The elastic modulus is modeled as a random variable with gamma distribution, 
being the probability distribution of this parameter obtained by the use of
the maximum entropy principle. 

An analysis of the model is performed, indexed by a dimensionless parameter
which describes the ratio between the discrete/continuous mass of the system.
This analysis shows that the dynamics of the random system is significantly 
altered when the values of the lumped mass are varied.
It is observed that this system right extreme behaves, in the limiting case where 
the lumped mass is very large, such as a mass-spring system. Also, one can note 
an irregular distribution of energy through the spectrum of frequencies, 
maybe induced by the cubic nonlinearity. Furthermore, the probability distributions 
of the lumped mass velocity present asymmetries and multimodal behavior, being 
this multimodality associated with the existence of areas of greater probability 
for the dynamic system response.


\section*{Acknowledgments}

The authors are indebted to the Brazilian agencies CNPq, CAPES, and FAPERJ 
for the financial support given to this research. They also wish to thank the 
anonymous referees', for useful comments and suggestions.

\section*{References}



\end{document}